\documentclass[useAMS, usenatbib, usegraphicx]{mn2e}
\title[Debris from terrestrial planet formation]{Debris from terrestrial planet formation: the Moon-forming collision}
\author[A.P. Jackson and M.C. Wyatt]{Alan P. Jackson$^{1}$\thanks{E-mail: ajackson@ast.cam.ac.uk (APJ);\newline wyatt@ast.cam.ac.uk (MCW)} and Mark C. Wyatt$^1$\footnotemark[1]\\%
$^1$Institute of Astronomy, University of Cambridge, Madingley Road, Cambridge, CB3 0HA}
\date{Submitted 2012}
\pagerange{\pageref{firstpage}--\pageref{lastpage}}
\pubyear{2012}
\usepackage[total={17.8cm,24.0cm},centering]{geometry}
\usepackage{times}
\usepackage[fleqn]{amsmath}
\usepackage[none]{hyphenat}
\usepackage{color}

\begin{document}
\label{firstpage}
\maketitle
\begin{abstract}
We study the evolution of debris created in the giant impacts expected during the final stages of terrestrial planet formation.  The starting point is the debris created in a simulation of the Moon-forming impact.  The dynamical evolution is followed for 10~Myr including the effects of Earth, Venus, Mars and Jupiter.  The spatial distribution evolves from a clump in the first few months to an asymmetric ring for the first 10~kyr and finally becoming an axisymmetric ring by about 1~Myr after the impact.  By 10~Myr after the impact 20\% of the particles have been accreted onto Earth and 17\% onto Venus, with 8\% ejected by Jupiter and other bodies playing minor roles.  However, the fate of the debris also depends strongly on how fast it is collisionally depleted, which depends on the poorly constrained size distribution of the impact debris.  Assuming that the debris is made up of 30\% by mass mm-cm-sized vapour condensates and 70\% boulders up to 500~km, we find that the condensates deplete rapidly on $\sim$1000~yr timescales, whereas the boulders deplete predominantly dynamically.  By considering the luminosity of dust produced in collisions within the boulder-debris distribution we find that the Moon-forming impact would have been readily detectable around other stars in Spitzer 24~$\mu$m surveys for around 25~Myr after the impact, with levels of emission comparable to many known hot dust systems.  The vapour condensates meanwhile produce a short-lived, optically thick, spike of emission.  We use these surveys to make an estimate of the fraction of stars that form terrestrial planets, $F_{\rm{TPF}}$.  Since current terrestrial planet formation models invoke multiple giant impacts, the low fraction of 10-100~Myr stars found to have warm ($\ga$150~K) dust implies that $F_{\rm{TPF}}\la$10\%.  For this number to be higher, it would require that either terrestrial planets are largely fully formed when the protoplanetary disk disperses, or that impact generated debris consists purely of sub-km objects such that its signature is short-lived.
\end{abstract}
\begin{keywords}
planetary systems:formation -- planets and satellites: formation -- Earth -- Moon
\end{keywords}

\section{Introduction}
\label{intro}
The current leading theory for the formation of the Moon is that of a low velocity, glancing, giant impact between a roughly Mars-sized body (named `Theia' by \citealt{halliday2000}) and the proto-Earth in the late stages of its formation.  See \citet{canup2004b} and \citet{canup2008} for a review and recent refinements.  Such a giant impact would produce a large quantity of debris that would escape the Earth entirely as well as that which would remain bound in Earth orbit and eventually accumulate to form the Moon.  While the precise quantity of debris produced, and in particular the amount of material placed into bound planetary orbit and so the potential for the formation of a massive satellite, depends on the collision parameters and properties of the colliding bodies (\citealt{canup2004a, agnor2004, canup2008}), the production of debris is quite generic.

Giant impacts are predicted to have been common during the final, chaotic growth, stage of terrestrial planet formation (e.g. \citealt{kenyon2006, raymond2009, kokubo2010}) with the terrestrial planets being built up sequentially through a series of giant impacts.  As well as the origin of the Moon giant impacts have also been suggested to explain the large core fraction of Mercury (\citealt{anic2006, benz2007}), the Martian hemispheric dichotomy (\citealt{wilhelms1984}; \citealt*{andrews-hanna2008, marinova2008}; \citealt{melosh2008, nimmo2008}) and the origin of the Pluto-Charon system (\citealt{canup2005, stern2006}).

As mentioned above the production of debris is a generic feature of giant impacts.  Indeed the models of the formation of Mercury by \citet{anic2006} and \citet{benz2007} explicitly rely on the expulsion of a large fraction of the mass of the colliding objects as debris.  Simulations of the chaotic growth phase of terrestrial planet formation do not follow the evolution of impact-produced debris in a completely consistent way however.  Earlier simulations such as \citet{chambers2004} and \citet{kenyon2006} typically used a perfect-merger model for the outcome of collisions, but detailed smoothed-particle hydrodynamic (\textsc{sph}) modelling of individual collisions such as that of \citet{agnor2004} and \citet*{asphaug2006} showed that such a perfect-merger model is not realistic.  In particular the class of `hit-and-run' collisions, in which the bodies collide without significant accretion or disruption, have been identified as important.  Incorporating a more realistic accretion model into N-body simulations has proved exceptionally computationally difficult however.  \citet{kokubo2010} used a slightly more realistic model by defining a critical impact velocity to separate collisions into perfect-merging and elastic hit-and-run regimes.  \citet{genda2011} present probably the most ambitious study to date utilising a hybrid N-body code which spawns an \textsc{sph} simulation when a collision occurs to track the outcome.  While this is the first full N-body simulation to estimate the amount of debris produced during the terrestrial planet formation process (the Monte-Carlo study of \citealt{stewart2012} produces similar results), in common with past studies they cannot follow the evolution of the debris.

Debris produced during terrestrial planet formation, its evolution and final destination, is of potential interest for several reasons.  Many young stars are observed to host disks of debris from either infra-red excesses or direct imaging (see e.g. \citealt{wyatt2008} for a review).  Such debris has been suggested as a possible signpost of the formation of terrestrial planets, both cold dust as an indicator of a dynamically stable environment (e.g. \citealt{raymond2011}) and hot dust as a direct signature from giant impacts (e.g. \citealt{lisse2008, lisse2009}).  The presence of debris in the system also has the potential to damp the orbits of the larger terrestrial planets and solve the problem that the terrestrial planets produced by current N-body simulations tend to be on orbits that are too dynamically excited (e.g. \citealt{obrien2006}).

Although debris has been suggested as a possible signpost of terrestrial planet formation, and a few specific debris systems have been linked with giant impacts, there has, as yet, been no study of the evolution of debris produced in terrestrial planet formation.  With this work we seek to begin filling in this area.  In particular there are a number of questions that need to be answered in relation to debris produced in giant impacts during terrestrial planet formation, such as: What happens to the debris, where does it go and what processes control its evolution?  Is the debris detectable, and over what timescales?  Can we use observations of dusty systems to constrain the frequency of terrestrial planet formation?

In this work we investigate the evolution of debris produced after a single giant impact event in an otherwise stable system.  Specifically we investigate the Moon-forming impact, which at $\sim$50~Myr after the formation of the solar system is thought to be the last giant impact to occur in the inner solar system.  The Moon-forming impact is the best studied example of a giant impact, aided by the constraints placed by the existence and properties of the Moon and its orbit, making it a logical starting point for the study of giant impact debris.

We first set out the configuration of the debris immediately after the Moon-forming impact in Section~\ref{initconds}.  Then in Section~\ref{dynamics} we present N-body simulations that follow the dynamical evolution of the debris over time, initially considering interactions with Earth only, and then introducing the effects of the other solar system planets.  In Section~\ref{collisions} we move on to discuss the other major evolutionary process influencing the debris, namely mutual collisions between debris bodies which result in a gradual grinding down of objects in the debris disk. In Section~\ref{comparison} we then compare the predicted observational signature of the debris produced by the Moon-forming impact with known debris disks, and use this to place constraints on the fraction of stars that undergo terrestrial planet formation.  Finally in Section~\ref{compositions} we discuss the potential implications of accretion of substantial amounts of giant impact debris for the solar system planets, and for terrestrial planet formation in general, before summarising our results in Section~\ref{conclusions}.

\section{Initial conditions}
\label{initconds}

We obtain the initial spatial and velocity distributions of the giant impact produced debris from an \textsc{sph} simulation of the Moon-forming impact which can be found at the top of Table 1 in \citet{marcus2009}.  This simulation was conducted using a modified version of the \textsc{gadget sph} code\footnote{see \citet{springel2005}} with total system mass $1.06 M_{\oplus}$ and impactor to target mass ratio 0.13 with collision velocity equal to the mutual escape velocity.  From this simulation those \textsc{sph} particles escaping the Earth-Moon system at 1 day after impact (2073 particles in total) were extracted and used to determine the initial parameters for the resulting debris cloud.

Escaping material accounts for 1.6\% of the total system mass, which, rescaling the mass of the post-impact Earth to $1 M_{\oplus}$, amounts to just under $10^{23}$ kg or $\sim$ 1.3 Lunar masses ($M_{\rm{L}}$).  This is very similar to the mass of the proto-lunar disk.  Around 40\% of the escaping material derives from the proto-Earth while the reminder originates from the impacting body.

As would be expected there is some anisotropy in the distribution of escaping \textsc{sph} particles in the plane of impact.  Unfortunately it is not possible to determine the true orientation of the interaction plane with respect to the plane of Earth's orbit. Indeed \citet{kokubo2010} find the obliquity distribution of terrestrial planets at the end of their planet formation simulations to be isotropic, indicative of an isotropic distribution of interaction planes in giant impacts.  With this lack of information about the true system orientation we opt to model the event producing the escaping particles as an isotropic explosion.  This is achieved by collapsing the three Cartesian coordinates into a single radial coordinate and then assuming spherical symmetry.

\begin{figure}
\includegraphics[width=85mm]{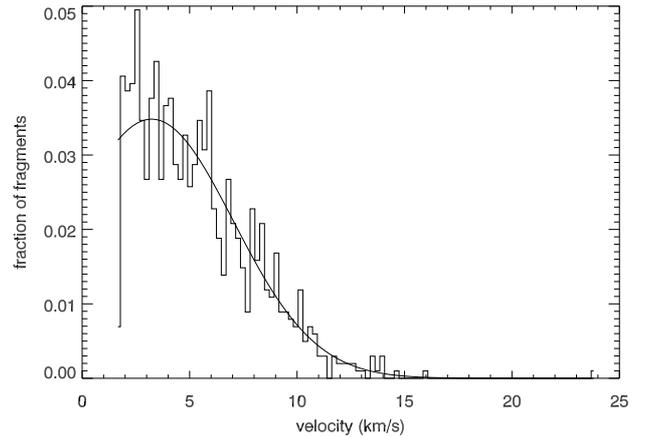}
\caption{Velocity distribution of the escaping fragments at 1 day after impact as taken from the \textsc{sph} simulation of \citet{marcus2009}.  Also shown is our fit to the distribution using a truncated Gaussian profile.}
\label{fragvel}
\end{figure}

The velocities of escaping \textsc{sph} particles are largely radial, with median deviation from purely radial of only 8.3 degrees, so we consider that it is reasonable to approximate the event as a purely radial explosion.  In addition, as above, we cannot determine the orientation of any anisotropies in the velocity distribution of escaping particles relative to the plane of Earth's orbit.  The majority of the escaping \textsc{sph} particles occur as part of small bound associations (fragments) but there are no obvious differences in the position or velocity distributions of fragments of different masses.

The best fit to the radial velocity distribution (see Fig.~\ref{fragvel}) of the escaping fragments is a truncated Gaussian with mean 3.23~km~s$^{-1}$, standard deviation 3.81~km~s$^{-1}$, and truncated below 1.67~km~s$^{-1}$.  This is preferred over a power law distribution as it provides a better fit to the high velocity tail.  The escaping \textsc{sph} particles have a mean radial position of ~50 Earth radii from the centre of mass (at 1 day after impact) and as such we choose to start our model using the truncated Gaussian velocity distribution above and with all particles at 50 Earth radii ($R_{\oplus}$) from the centre of mass.  The truncation of the distribution is just above the escape velocity at 50~$R_{\oplus}$ of 1.58~km~s$^{-1}$, which ensures that all of the particles will escape.  We take the pre-impact Earth to be on a circular orbit at 1~AU.  The mean velocity of particles generated according to this distribution is around 5.3~km~s$^{-1}$, and is roughly equal to the mean velocity at infinity (in the absence of the Sun), due to the small escape velocity at this distance.

\subsection{Orbits of the debris}
\label{debrisorb}

Since the debris material is thrown off Earth the orbits of the debris particles can be determined assuming their velocities to be that of Earth plus a kick in velocity, as defined below.  The orbit of Earth is initially very close to circular and without loss of generality we can define our Cartesian reference such that the orbit of Earth is in the $x-y$ plane with Earth lying on the $x$-axis at the time of impact for convenience.  The semi-major axis, eccentricity and inclination of the new orbits of the debris ($a'$, $e'$, $I'$) will be given by:
\begin{equation}
\label{aaprim}
\frac{a}{a'}=1-2\left(\frac{\Delta v}{v_{\rm{k}}}\right) S_{\theta}S_{\phi}-\left(\frac{\Delta v}{v_{\rm{k}}} \right)^2,
\end{equation}
\begin{equation}
\label{eprim2}
e'^2=1-\left(\frac{h'^2}{h^2}\frac{a}{a'}\right),
\end{equation}
\begin{equation}
\label{Iprim}
\sec(I')=\left[1+\left(\frac{\left(\frac{\Delta v}{v_k}\right)C_{\theta}}{1+\left(\frac{\Delta v}{v_k}\right)S_{\theta}S_{\phi}}\right)^2\right]^{\frac{1}{2}},
\end{equation}
where,
\begin{equation}
\label{hhprim}
\frac{h'^2}{h^2}=1+2\left(\frac{\Delta v}{v_k}\right)S_{\theta}S_{\phi}+\left(\frac{\Delta v}{v_k}\right)^2(C_{\theta}^2+S_{\theta}^2S_{\phi}^2),
\end{equation}
\begin{equation}
\label{vk}
v_{\rm{k}}^2=\frac{GM_{\odot}}{a},
\end{equation}
and we use the common shorthand notation, $S_x$ and $C_x$ for $\sin(x)$ and $\cos(x)$.  The variables $\theta$ and $\phi$ are spherical polar angles centred at Earth describing the direction of the kick relative to the $X$-axis and $x-y$ plane, $\Delta v$ is the magnitude of the kick and $a$ is the semi-major axis of the old orbit (1~AU).  These equations, as well as more general forms and applications are described in Jackson et al. (2012, in prep).

\begin{figure}
\includegraphics[width=80mm]{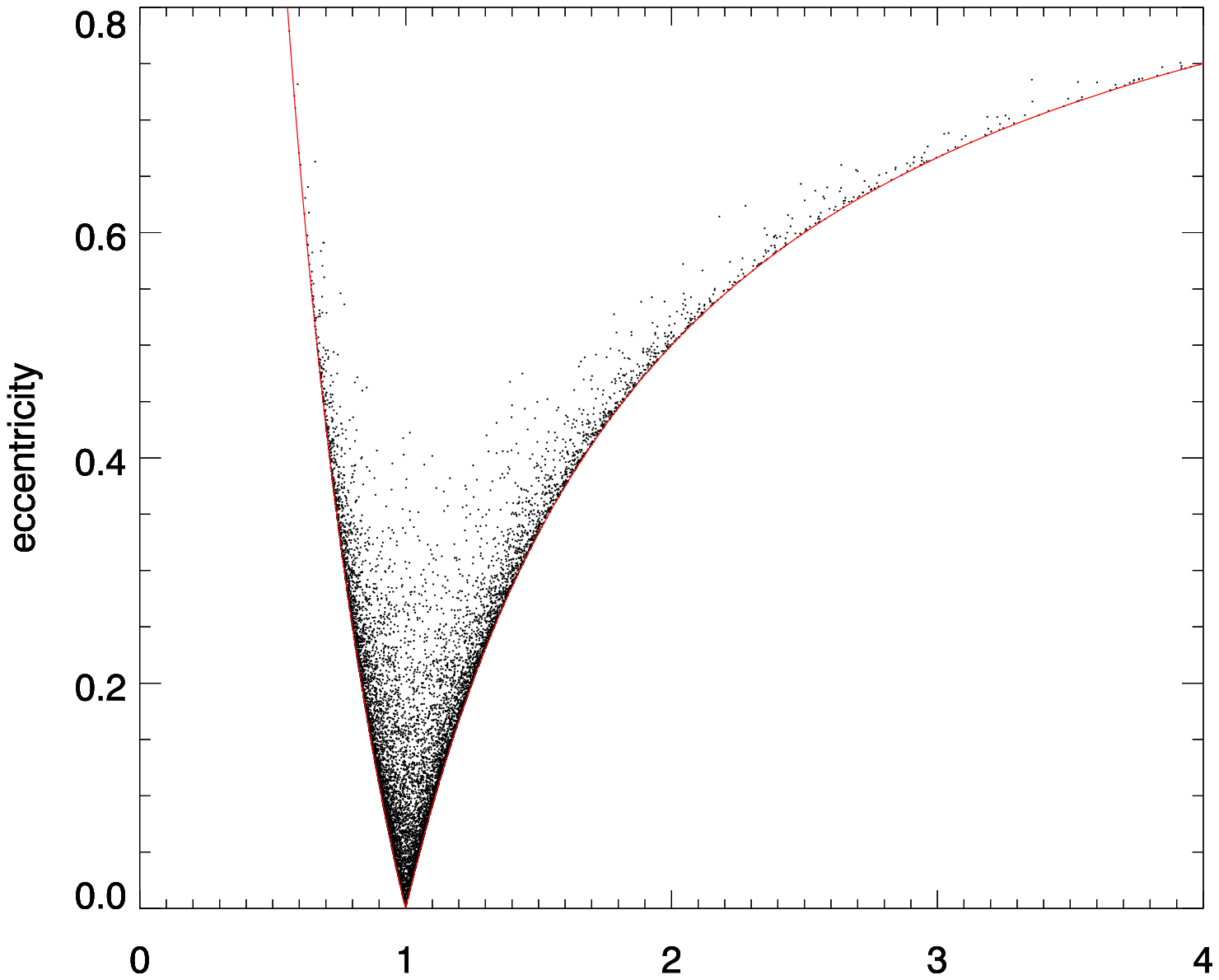}
\includegraphics[width=80mm]{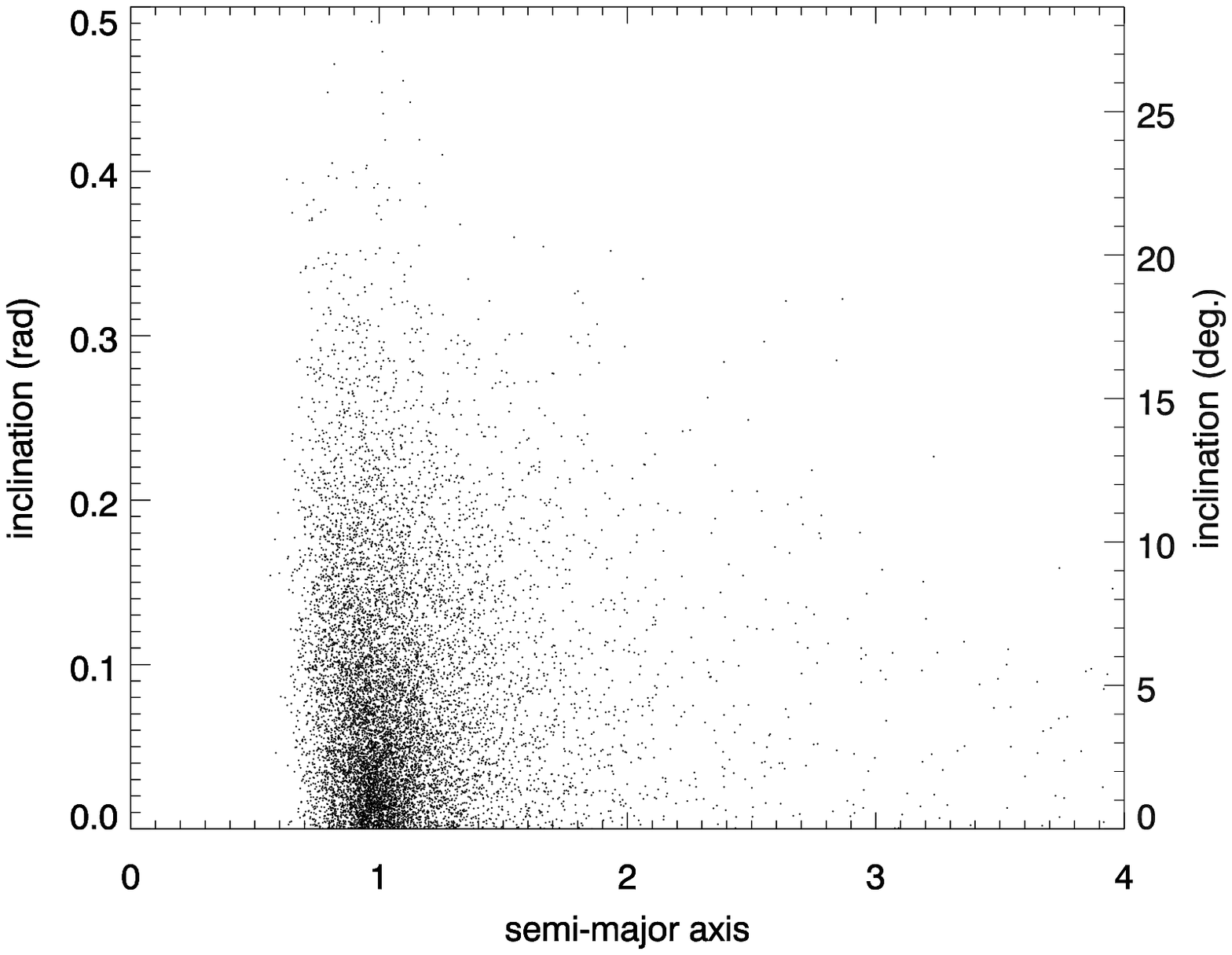}
\caption{The eccentricity -- semi-major axis and inclination -- semi-major axis distributions for a set of 10000 particles generated according to our fit to the velocity distribution of ejected fragments.  The red lines on the upper plot indicate the apocentre and pericentre conditions on the orbits described in the text.}
\label{vrcomp}
\end{figure}

To determine the exact $\Delta v$ for these equations we must take into account the energy needed to escape Earth, so $\Delta v = \sqrt{v_{\rm{p}}^2-v_{\rm{esc}}^2}$ where $v_{\rm{p}}$ is the intial particle velocity and $v_{\rm{esc}}$ is the escape velocity at 50~$R_{\oplus}$, for the distribution described above.  Since the escape velocity at 50~$R_{\oplus}$ is rather small the difference between $\Delta v$ and the particle velocities for the distribution described above is also quite small.  In combination with the truncated Gaussian velocity distribution and an isotropic explosion these equations produce an eccentricity -- semi-major axis and inclination -- semi-major axis distribution for the debris as shown in Fig.~\ref{vrcomp}.  This distribution is readily explained by the fact that the debris orbits must pass through the point at which the velocity kick is applied, i.e. the point at which the Moon-forming collision occurs.  This is a fixed point in space and we refer to it as the `collision point'.  For example, this leads to a constraint on the orbits of the debris that no debris particle can have its apocentre at less than 1~AU and nor can it have its pericentre at greater than 1~AU.

\section{Dynamical evolution of the debris}
\label{dynamics}

\begin{figure*}
\includegraphics[width=55mm]{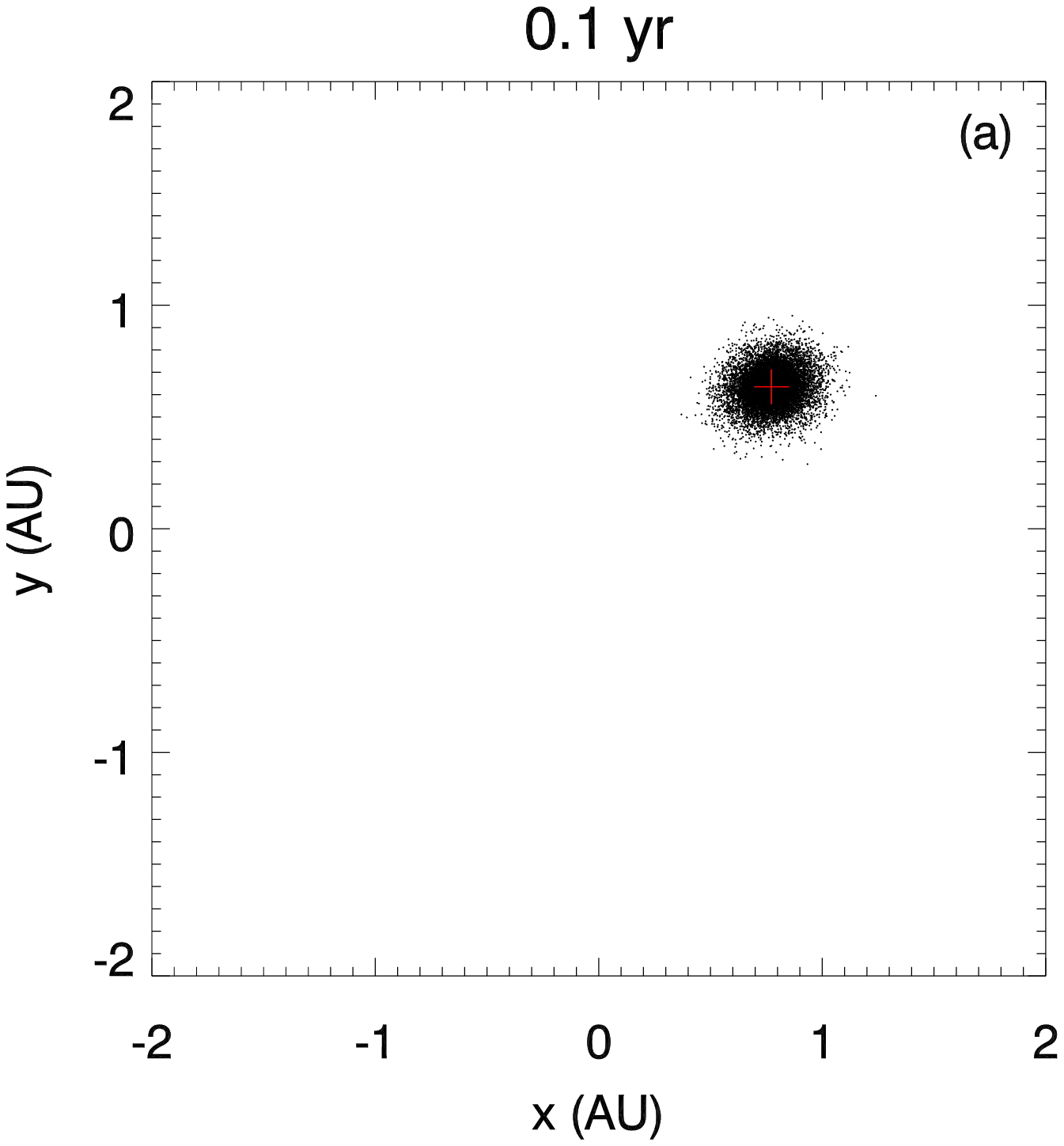}
\includegraphics[width=55mm]{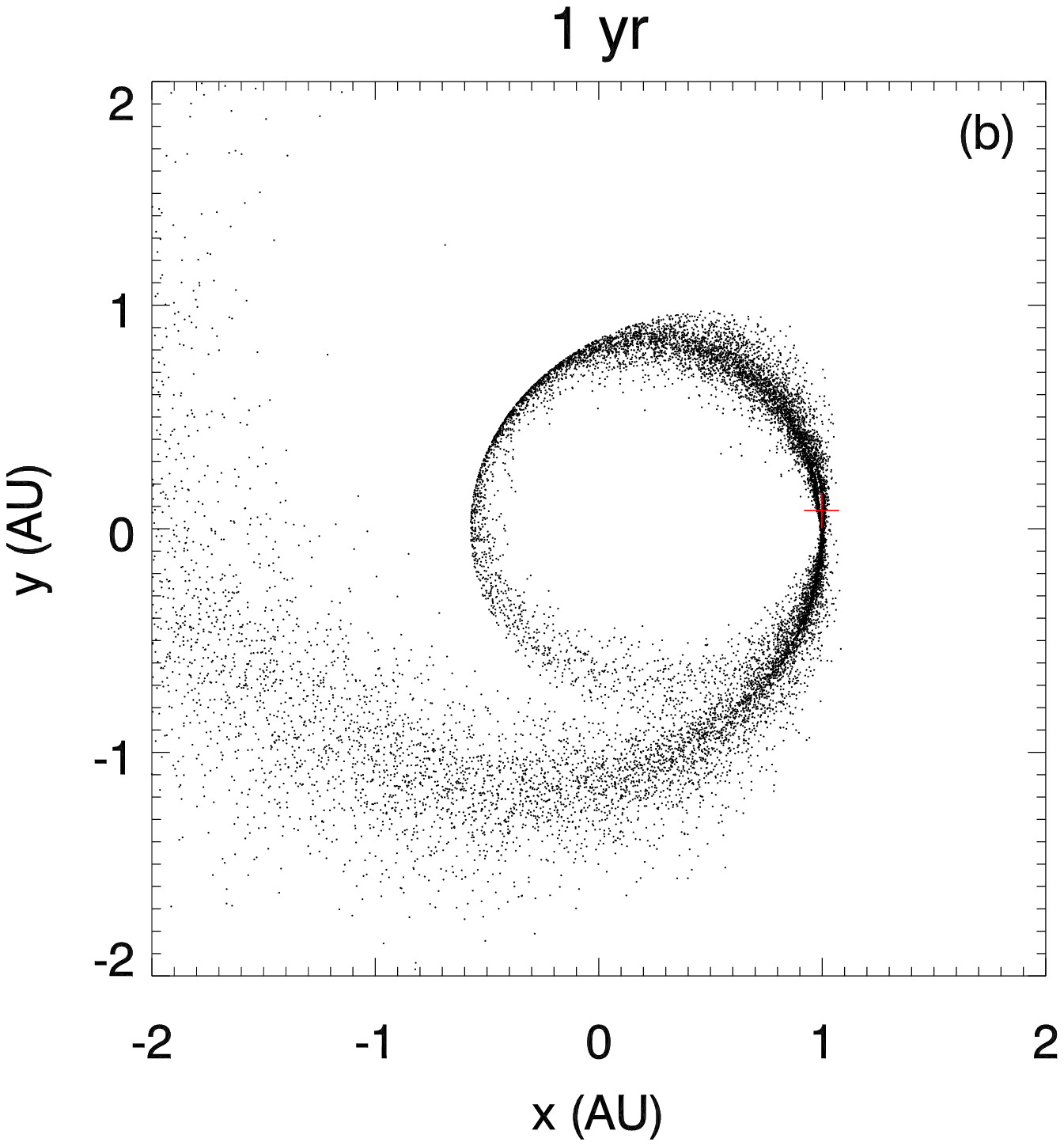}
\includegraphics[width=55mm]{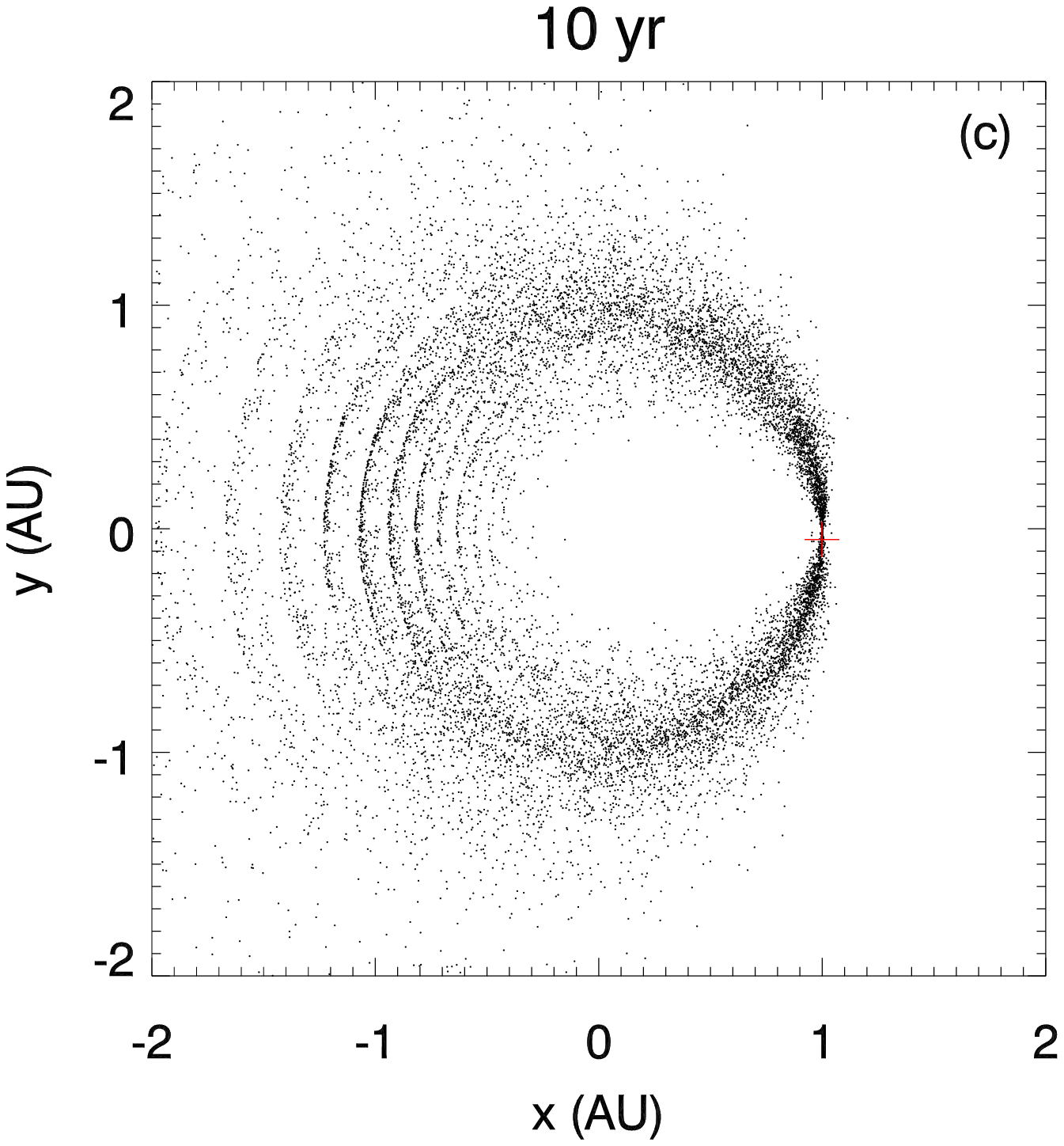}
\includegraphics[width=55mm]{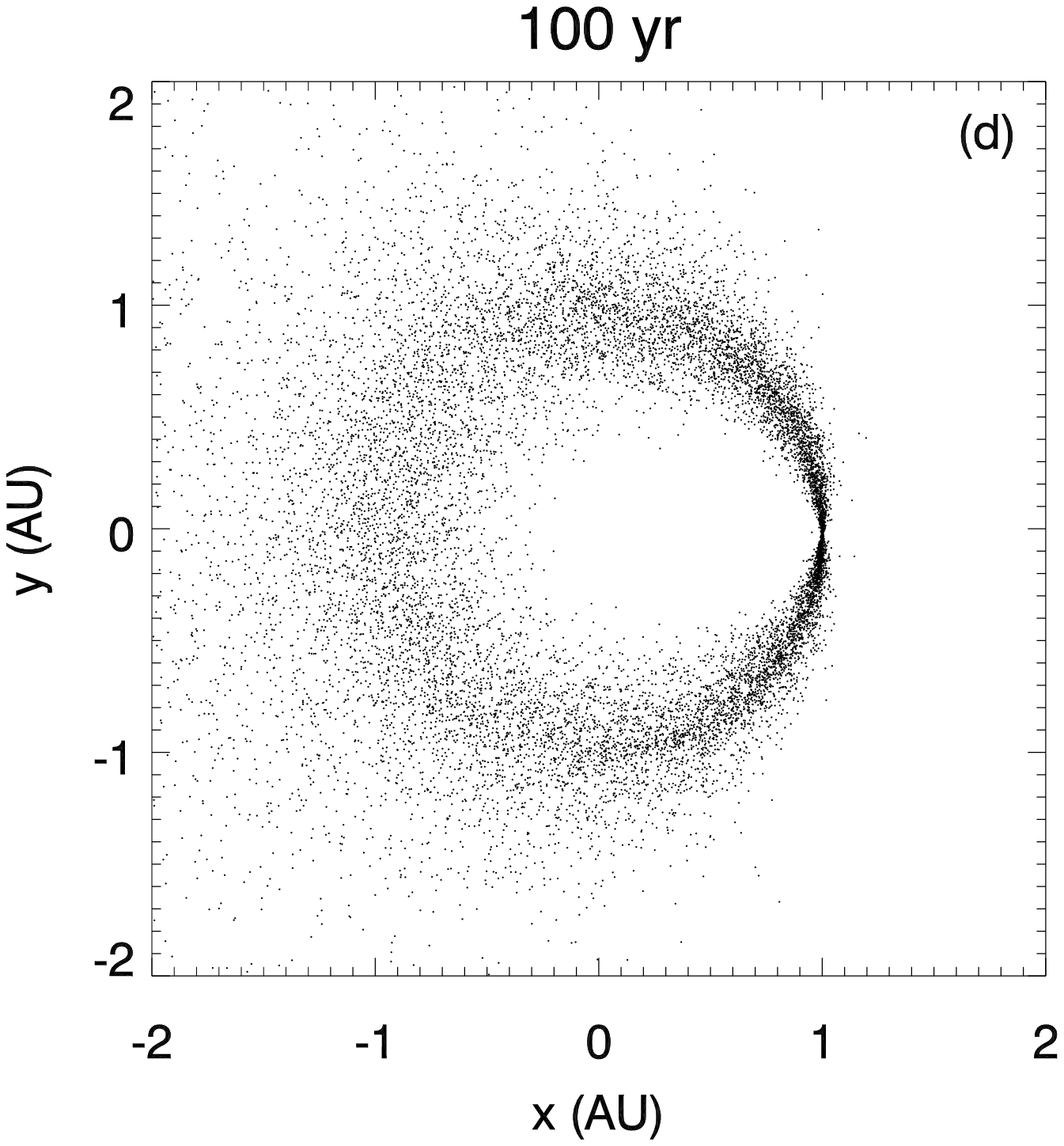}
\includegraphics[width=55mm]{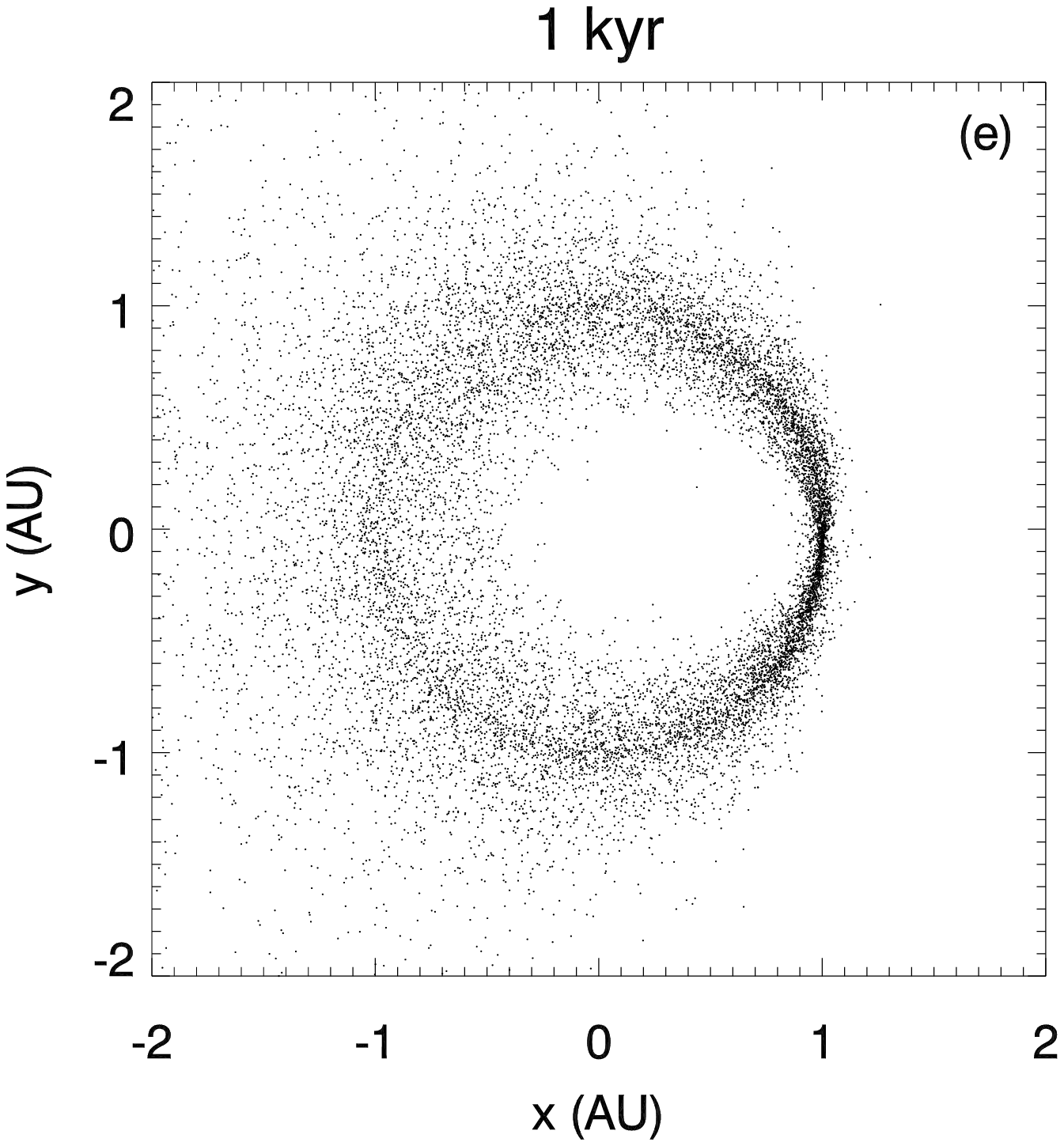}
\includegraphics[width=55mm]{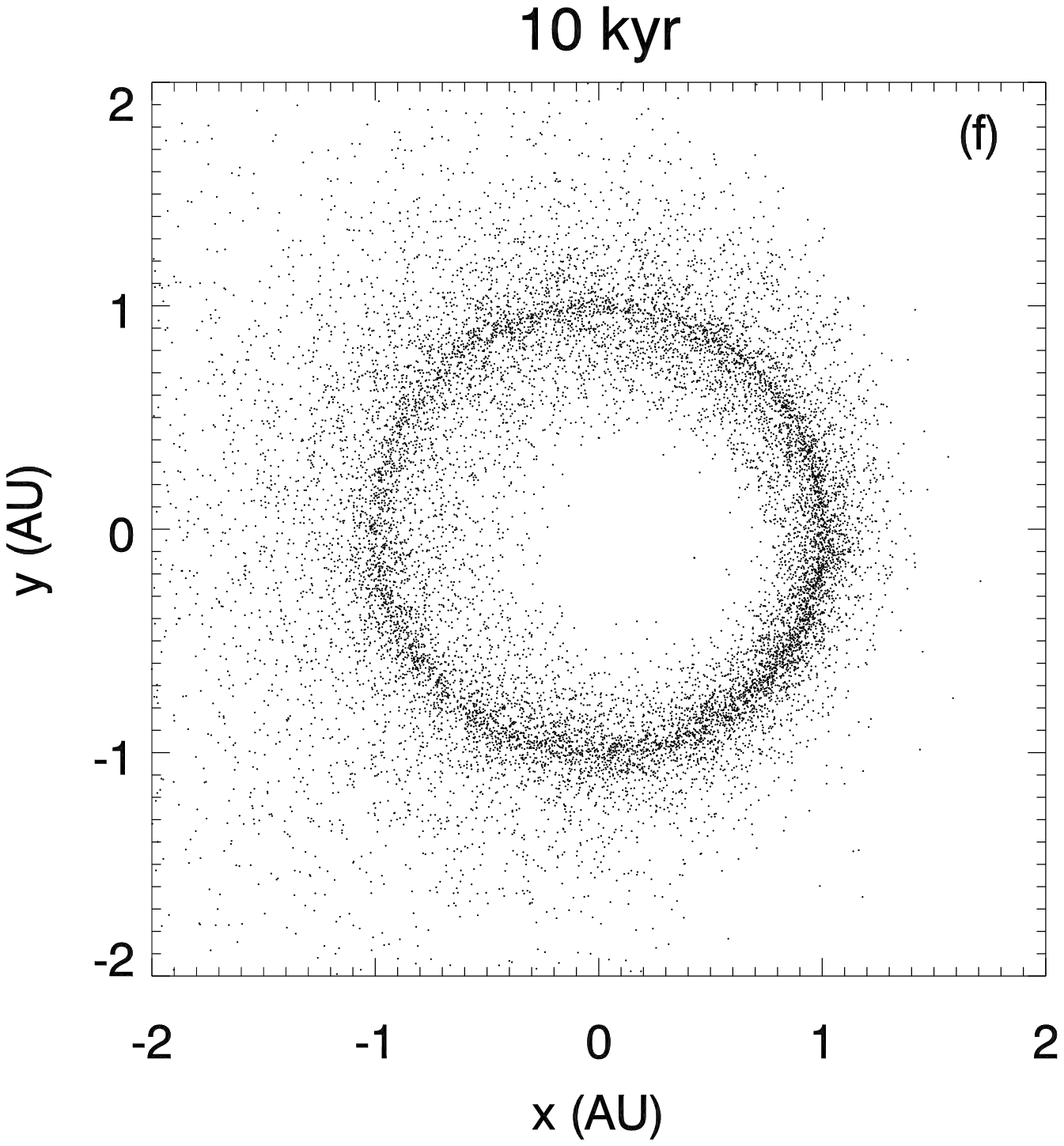}
\includegraphics[width=55mm]{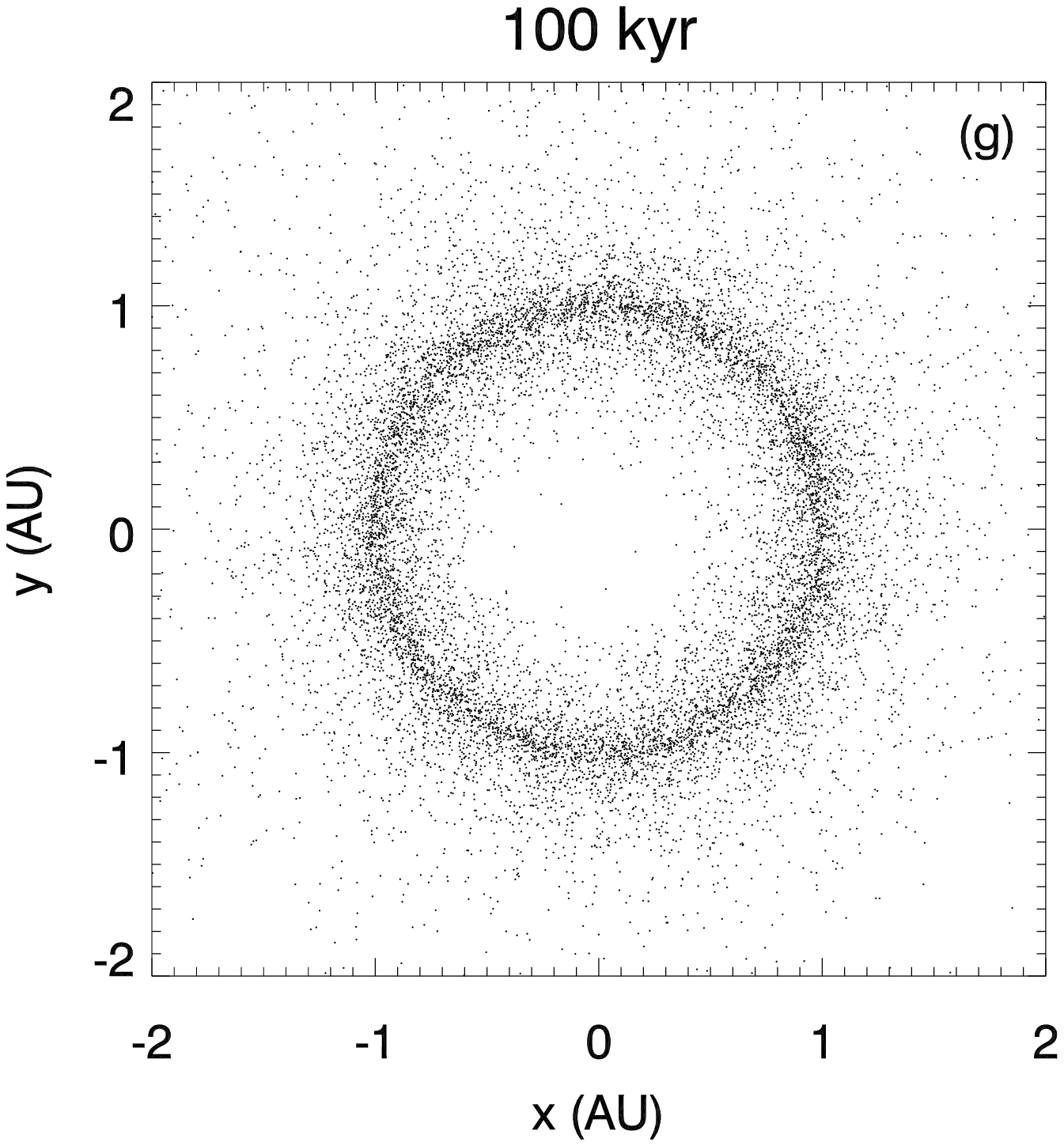}
\includegraphics[width=55mm]{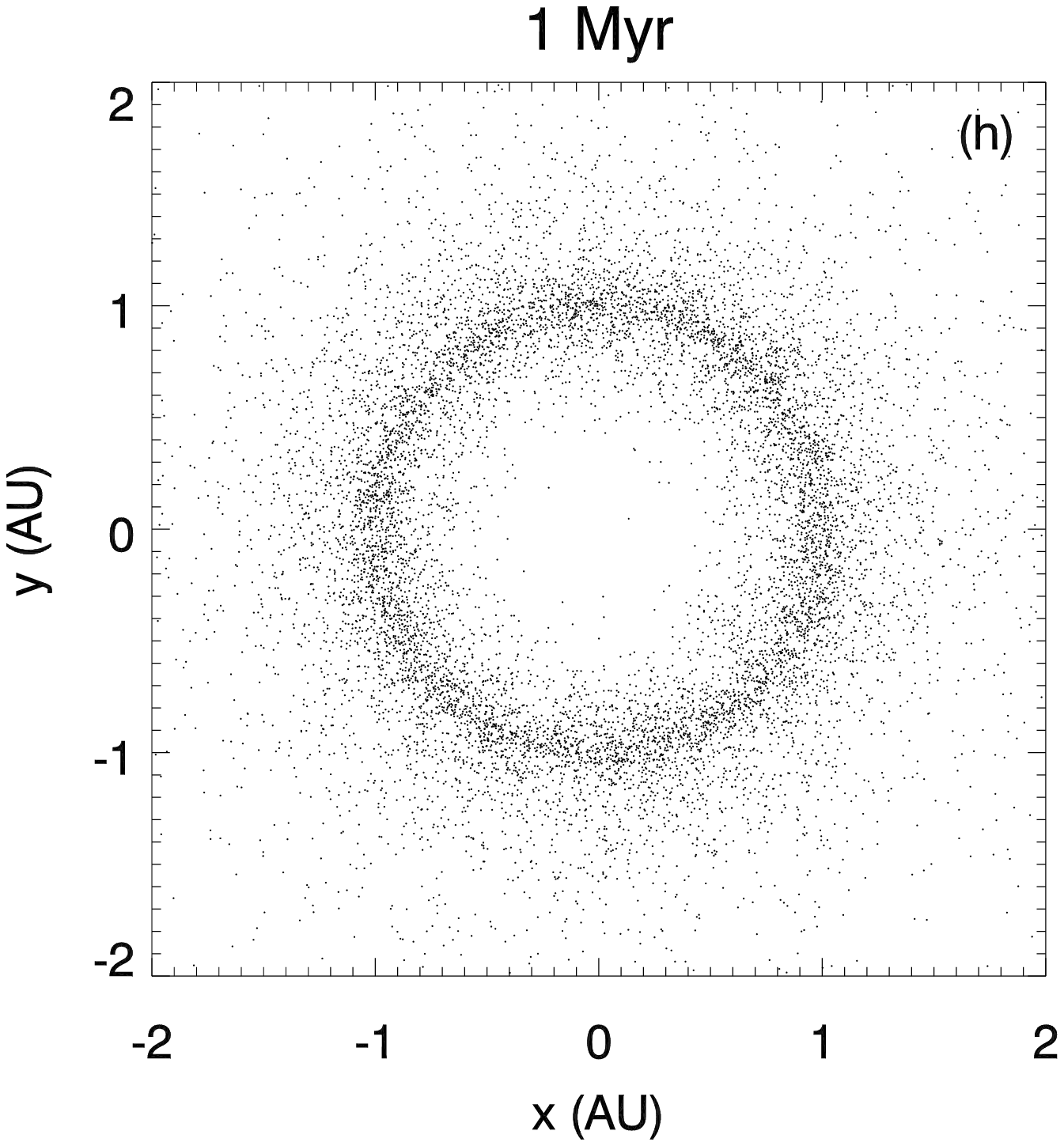}
\includegraphics[width=55mm]{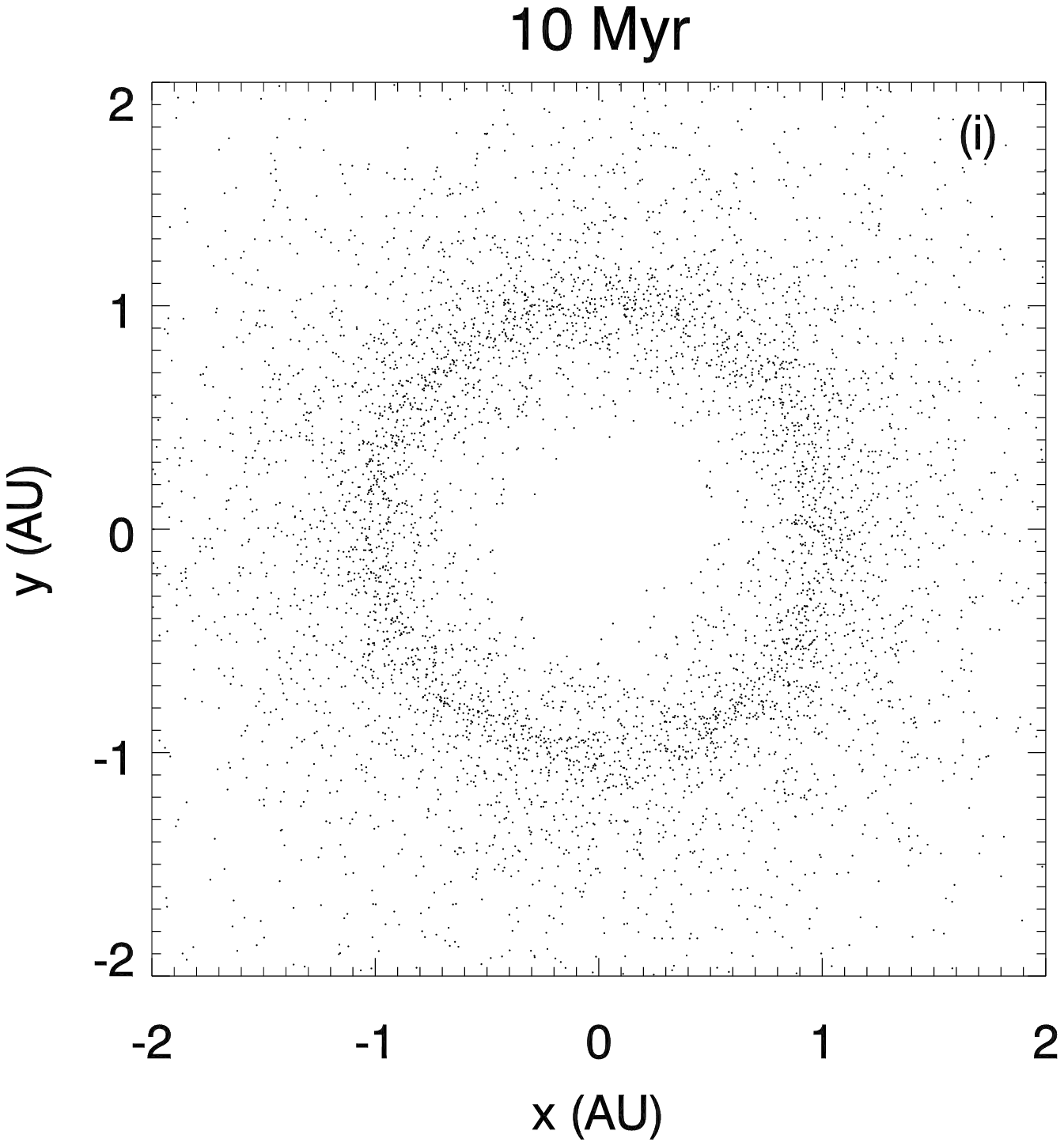}
\caption{Snapshots of the evolution of the debris ejected by the Moon-forming impact for the DE simulation.  The disk is viewed face on with the Sun at the origin and the collision point at the position (1,0).  In the upper row we also indicate the position of Earth with a red cross.  See Section \ref{DEoverview} for further information.}
\label{DEsimdisc}
\end{figure*}

In this Section we discuss the dynamical evolution of the debris over long timescales after its ejection in the initial impact.  For all of the simulations described below we use the hybrid integrator option of the \textsc{mercury} orbital integrator\footnote{see \citet{chambers1999}}.  In this mode particles are integrated symplectically when far from a massive body, switching to a Bulirsch-Stoer integrator to track close-encounters, with the switch-over distance set as 3 Hill radii.  Collisions with the massive bodies in the system are tracked and the debris is treated as massless point-like test particles.  When a debris particle is further than 100 AU from the Sun it is removed from the simulation and is considered to have been ejected.

\subsection{The Moon}
\label{lunarinfluence}

Immediately after the impact there will of course be no Moon, though there will be the (slightly more massive) proto-Lunar disk.  The Moon is believed to have coalesced relatively rapidly, however (\citealt*{ida1997, kokubo2000}), so for a large fraction of the time spanned by our simulations the Moon (or multiple smaller progenitors) would be present.  Nonetheless in all of our simulations we treat the Earth-Moon system as a single body of mass $1 M_{\oplus}$.

One of the primary considerations behind this is computational complexity, since being required to resolve the Earth-Moon orbital motion would place a significant additional burden on the simulations.  This is especially so since models of coalescence of the Moon from the proto-lunar disk (e.g. \citealt{ida1997, kokubo2000}) suggest that the Moon would have formed near the Roche limit at only a few $R_{\oplus}$.

The influence of the presence of the Moon on the evolution of the debris is also likely to be quite small, and thus neglecting it is unlikely to significantly alter our results.  \citet{bandermann1973} described analytically the ratio of accretion onto Earth and the Moon, $A_{\oplus}/A_{\rm{L}}$, as a function of the Earth-Moon separation and the relative velocity of the impacting meteoroids.  The minimum accretion ratio between Earth and the Moon is roughly the ratio of their geometrical cross-sections ($R_{\oplus}^2/R_{\rm{L}}^2$=13.4), but this limit is only approached as the relative velocity becomes very large (such that gravitational focussing is unimportant).  Although $A_{\oplus}/A_{\rm{L}}$ decreases as the Earth-Moon separation decreases, even for a separation of 3~$R_{\oplus}$ the typically $\sim$5~km~s$^{-1}$ relative velocities of the debris result in an accretion ratio of $\sim$25.  Tidal dissipation also raises the lunar orbit quite rapidly (\citealt{touma1994, zahnle2007}). Thus $A_{\oplus}/A_{\rm{L}}$ also rises, initially quite rapidly, before flattening out at $\sim$50 for large separations (and relative velocities of $\sim$5~km~s$^{-1}$).  As the greatest influence of Earth on the debris is through accretion, this suggests that the influence of the Moon would be small in comparison.

\subsection{Interactions with Earth only}
\label{DEdyn}

To understand the dynamical evolution of debris ejected by a giant impact in the context of a planetary system like the solar system, it is first important to understand the evolution of debris in the presence of the parent body (Earth) and host star alone.  Here we discuss the results of a 14000 particle simulation run for 10 Myr with only Earth (and Sun) present, hereafter referred to as the DE simulation.  The particles were set up with initial conditions as described above with the collision point located on the x-axis and then integrated forward using \textsc{mercury}.  Earth is placed on the x-axis at the centre of the debris cloud on a very nearly circular orbit ($e<10^{-4}$).

\subsubsection{Overview of the evolution}
\label{DEoverview}

Fig.~\ref{DEsimdisc} shows snapshots of the debris disk at various stages of evolution.  In the first year we see the debris spreading away from Earth and forming a whip-like spiral as material on orbits with periods less than one year advances ahead of Earth, while material on longer period orbits trails behind (Fig.~\ref{DEsimdisc}(b)).  By 10 years after impact this has produced a tightly wound spiral-like structure (Fig.~\ref{DEsimdisc}(c)).

After 100 years scattering by Earth during collision-point passages has started to have an effect (see Section \ref{peracc}).  Coupled with the spiral structure becoming increasingly tightly wound individual spiral sections are no longer discernible (Fig.~\ref{DEsimdisc}(d)) and the disk has become continuous, albeit strongly asymmetric with a pronounced pinch at the collision point.  This pinch in the disk is fixed in space at the location of the initial collision point.  It does not follow Earth's orbit because all particles originated from this point, but were put onto orbits with a range of periods.  Although their initial post-collision orbits are constrained to pass through the collision point, the range of orbital periods mean that individual particles, and Earth, will generally pass through the collision point at different times.

The asymmetry is still very much present at 1~kyr (Fig.~\ref{DEsimdisc}(e)) and only fades after 10~kyrs (Fig.~\ref{DEsimdisc}(f)).  At such a time precession of particle orbits has symmetrised the region nearest to the orbit of Earth, though the regions interior and exterior to this remain asymmetric.  It is not until around 100~kyrs after impact (Fig.~\ref{DEsimdisc}(g)) that the disk attains a fully axisymmetric structure.  Once axisymmetry is attained the azimuthal structure of the disk is quite stable.  The evolution from 100~kyrs to 1~Myr (Fig.~\ref{DEsimdisc}(h)) and to 10~Myr (Fig.~\ref{DEsimdisc}(i)) is a gradual decrease in the surface density of the disk.  As a result of the geometry of the original explosion there is a tendency for particles very close to Earth's orbit, and thus particles undergoing interactions with Earth, to be near pericentre or apocentre.  This is true even at very late times, despite the fact that the disk becomes axisymmetric relatively early.

\begin{figure}
\includegraphics[width=85mm]{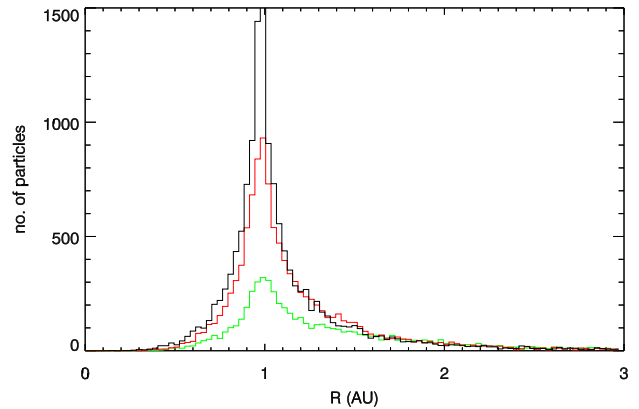}
\caption{Evolution of the radial density profile of the disk for the DE simulation.  Snapshots shown at 100 kyrs (black), 1 Myr (red) and 10 Myr (green).  The density is strongly peaked at 1 AU at early times with the 'peakedness' decreasing at later times as particles are re-accreted.}
\label{DEradhist}
\end{figure}

Aside from inducing symmetrisation on a timescale of 10~kyr, presumably through secular interactions, the influence of Earth on the large-scale structure of the disk is primarily through re-accretion of particles.  This will be discussed in more detail in Sections \ref{peracc} and \ref{Analacc} but is evident in Fig.~\ref{DEradhist} as the decrease in amplitude of the peak in density at 1 AU over time.

\begin{figure*}
\includegraphics[width=80mm]{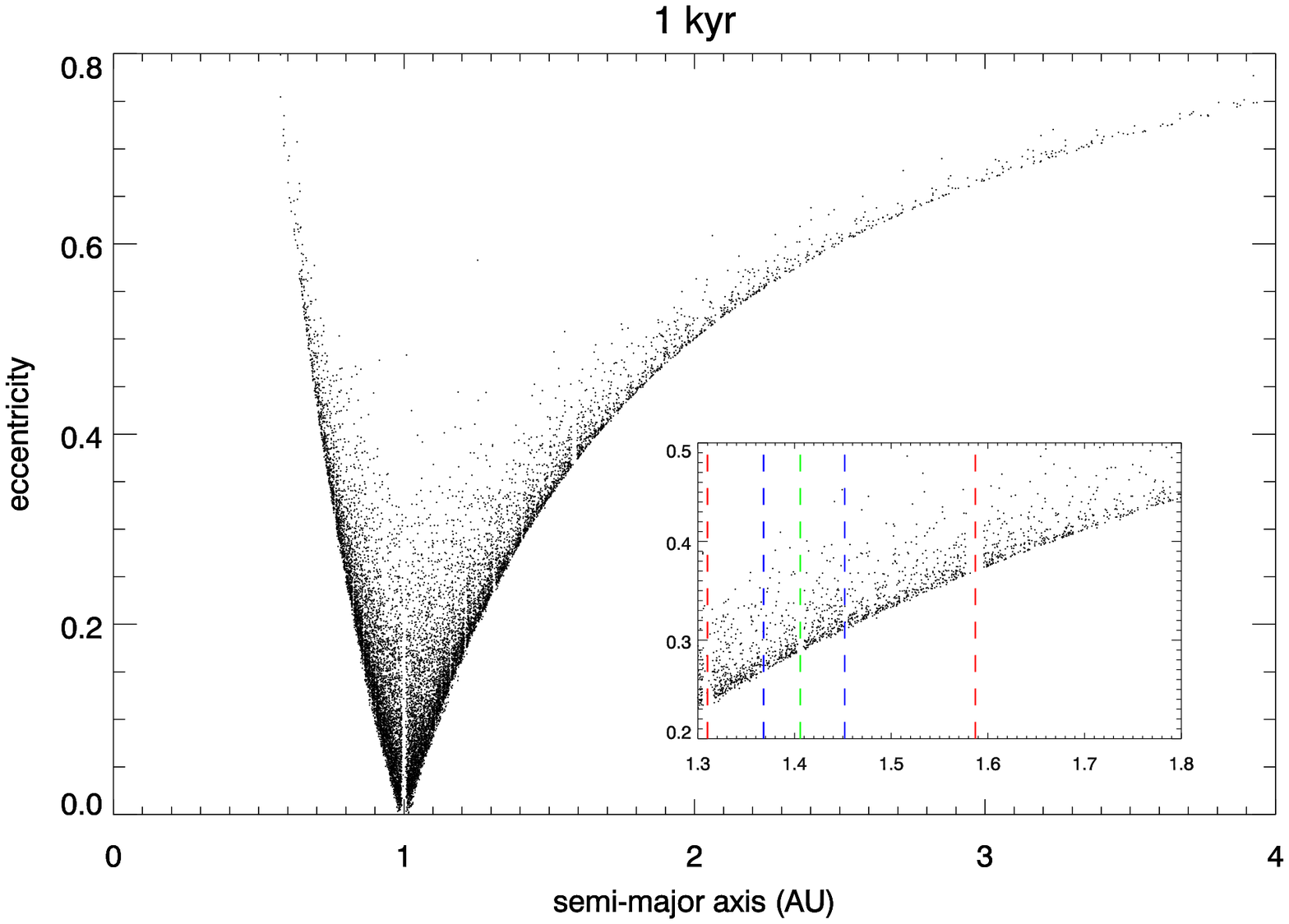}
\includegraphics[width=80mm]{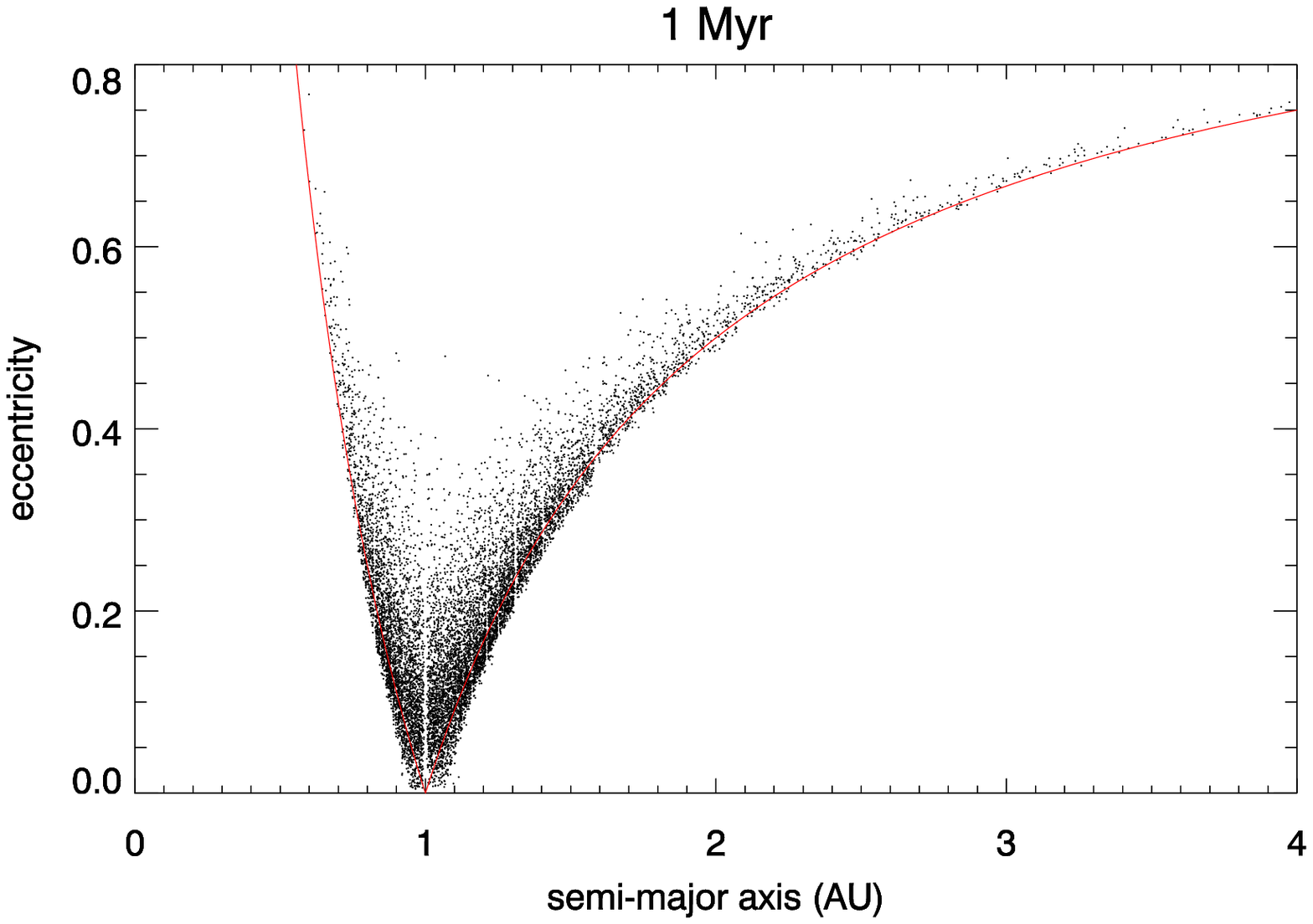}
\caption{Snapshots of the distribution of semi-major axes and eccentricities of the debris particles for the DE simulation at 1~kyr and 1~Myr.  Compare with the initial distribution in the upper panel of Fig.~\ref{vrcomp}.  The primary evolution is a decrease in density of particles, with the decrease preferentially affecting particles with semi-major axes near 1~AU.  The shape of the distribution remains largely unchanged as indicated by the red lines corresponding to the original outer boundary of the distribution on the 1~Myr plot.  The slight diffusion beyond the original boundaries is due to distant encounters.  Many narrow gaps in semi-major axis at the location of period commensurabilities with Earth are also visible, the most prominent being the 1:1.  As an inset to the 1~kyr plot we show an enlargement of one section to enhance the visibility of the gaps and mark the strongest ones with dashed lines, red for the 2:1 and 3:2, green for the 5:3 and blue for the 8:5 and 7:4.}
\label{DEaedist}
\end{figure*}

Although particles frequently undergo scattering encounters with Earth this does not induce significant changes in the broad structure, since particles tend to encounter Earth at similar velocities to that which was imparted to them by the impact.  In addition, since particles cannot be put onto orbits with apocentres less than, or pericentres greater than, 1~AU the distribution of semi-major axes and eccentricities changes little between 1~kyr and 1~Myr (see Fig.~\ref{DEaedist}). The distribution does spread marginally beyond the original boundaries of the initial distribution (red lines on right hand plot) however, because particles can still be scattered (albeit very weakly) even if their orbits do not cross that of Earth.

At the end of 10~Myr Earth has re-accreted 8081 out of 14000 particles, 58\% of the material.  The number of particles remaining in the disk, $N$, is reasonably well fit by a function of the form $N = N_0 \exp(-\sqrt{t/\tau})$, where the characteristic timescale, $\tau \sim 12$~Myr.  Very few particles are ejected from the system, with the majority of ejected particles being those that were put onto initial orbits with $e>1$.

\subsubsection{Periodic scattering and accretion}
\label{peracc}

\begin{figure}
\begin{center}
\includegraphics[width=55mm]{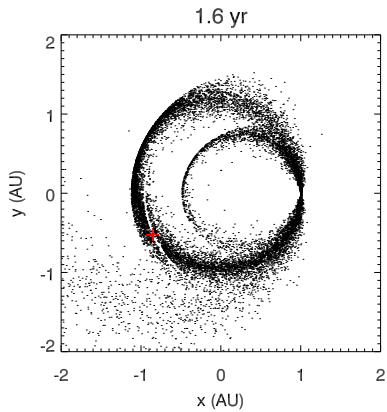}
\end{center}
\caption{Snapshot of the disk at 1.6 years after impact.  The slash through the disk near the position of Earth corresponds to particles that would have occupied the 1:1 period commensurability.  The range of eccentricities of the particles causes the gap to be point-like at the collision point, but spread into an arc further around the orbit.}
\label{discgap}
\end{figure}

The narrow gaps in Fig.~\ref{DEaedist} at the locations of period commensurabilities between Earth and the debris are not a product of typical resonant interactions.  Rather these gaps are the product of the initial conditions that all debris starts (essentially) at the position of Earth and that all orbits must pass through the collision point.  This means that debris particles on orbits with integer period commensurabilities with Earth will, at early times, pass through the collision point \emph{at the same time as Earth} on a repeating cycle set by the order of the commensurability.  Hence the 1:1 commensurability gap appears at 1 year after impact while the 2:1 gap appears at 2 years after impact, the 3:1 and 3:2 gaps at 3 years after impact and so on.  Those particles passing through the collision point at the same time as Earth will undergo close encounters and be scattered or accreted, either removing them from the distribution entirely or moving them to an orbit with a different semi-major axis, leading to the gaps in the distribution.  Fig.~\ref{discgap} shows the spatial counterpart of the 1:1 commensurability gap after the first collision point passage.  The range of eccentricities and arguments of pericentre of the particles that would have occupied the gap means that when Earth is at the opposite side of its orbit from the collision point the gap appears as an arc shaped slash through the disk.

Some particles, particularly near the edges of a commensurability will not have a very close encounter during the first cycle of the commensurability.  The repeating nature of the commensurabilities, however, means that they will continue to have encounters with Earth, until either the cumulative effect of many comparatively distant close encounters is enough to move them out of the commensurability, or they finally do have a very close encounter and undergo a more significant orbital change.  The width of the gap is set by the length of the repeat cycle.  Hence, in the inset in Fig.~\ref{DEaedist}, the 5:3 gap is narrower than the 2:1 and 3:2 gaps and wider than the 8:5 and 7:4 gaps.  It is also of note that, as well as the collision point, all of the particles initially share the same line of nodes. This means there is also a line on the opposite side of the Sun from the collision point (the `anti-collision line') through which all of the particles pass.  This makes it possible for some particles to undergo close encounters at half-integer times and for particles near the 1:1 commensurability to undergo close encounters at 6 month intervals, reinforcing the strength of the 1:1 gap.

\begin{figure}
\includegraphics[width=85mm]{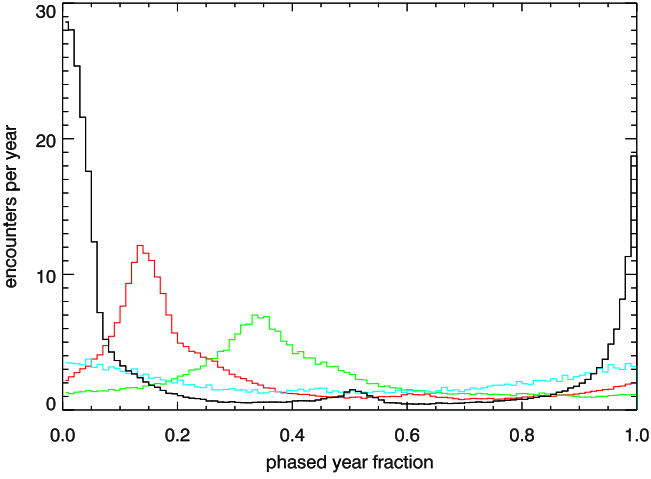}
\caption{Number of close encounters per year occurring between debris particles and Earth for the DE simulation, phase-folded on a timescale of 1 year.  The sum of the first kyr after impact is shown in black while red, green and cyan show 2-3~kyrs after impact, 5-6~kyrs after impact and 15-16~kyrs after impact respectively.  The strong spike in the encounter rate corresponds to Earth passing through narrowest part of the disk, initially at which Earth passes through the collision point, is clearly visible.  In the first kyr a smaller bump at phase 0.5 when Earth is passing through the anti-collision line is also discernible.  `Close encounters' are tracked out to closest approach distances of 0.03~AU (3 Hill Radii) from Earth.}
\label{DEphasenc}
\end{figure}

During the first thousand years re-accretion onto Earth occurs predominantly during passage of the collision point, and to a lesser extent during passage through the anti-collision line, and is thus strongly periodic.  This is demonstrated in Fig.~\ref{DEphasenc}, which shows the number of close encounters tracked by \textsc{mercury} phase-folded over one year.  Over time precession of the debris particle orbits shifts the narrowest point in the disk around Earth's orbit away from the original collision point and smears it out due to the different precession rates of different orbits.  The smearing effect of differing precession rates causes the periodicity of accretion to fade.  After around 15-20~kyr, the timescale for most particle orbits to complete a precession cycle, this is no longer a significant effect.

Although the periodic scattering and accretion is no longer a significant effect after the first 15-20~kyr, many of the gaps in the semi-major axis distribution at period commensurabilities remain open.  This is because if a particle slowly migrates, or is scattered, into one of the period commensurabilities, as its orbit precesses it will eventually begin to undergo close encounters with Earth, and will continue to do so until it is removed from the period commensurability or precession moves it away from encountering Earth.  For faster repeat cycle commensurabilities the timescale for a particle to be removed from the commensurability is shorter than the timescale for precession to move the particle away from undergoing close encounters.  Thus, although the scattering/accretion of particles at the commensurabilities is no longer synchronised, the gaps remain open.

\subsubsection{Analytic accretion rate}
\label{Analacc}

We can estimate the accretion rate directly from the equations describing the semi-major axis - eccentricity - inclination distribution from Section~\ref{debrisorb} if we assume a completely axisymmetric distribution.  If the total number of particles is $N$ and the number density of particles near Earth is $n$, then the accretion rate will be
\begin{equation}
R_{\rm{acc}}=-\frac{dN}{dt}=n \langle \sigma v_{\rm{rel}} \rangle,
\label{Raccav}
\end{equation}
where $\sigma$ is the cross-section for collision with Earth and $v_{rel}$ is the velocity of particles relative to Earth where they would be colliding.  As discussed earlier gravitational focussing is very important for re-accretion onto Earth as a result of the rather low velocities of the particles relative to Earth.  The $\sigma$ here is thus not simply the geometrical cross-sectional area of Earth but rather is significantly enlarged; incorporating the effect of gravitational focussing $\sigma$ is given by
\begin{equation}
\sigma=\pi R_{\oplus}^2\left(1+\left(\frac{v_{\rm{esc}}}{v_{\rm{rel}}}\right)^2\right),
\end{equation}
where $v_{\rm{esc}}$ is the escape velocity at Earth's surface, 11.2 km s$^{-1}$.  The addition of the gravitational focussing term enlarges the cross-section by a factor of $\sim$10.

An axisymmetric distribution will be roughly toroidal, centred at Earth's orbit.  The characteristic extent of the cross-section of this toroid in the radial direction is $a \langle e' \rangle$, while the characteristic extent in the $z$-direction is $a \langle I' \rangle$, where $\langle e' \rangle$ and $\langle I' \rangle$ are, respectively, the mean eccentricity and mean inclination of the debris.  The volume of this characteristic toroid is thus $8 \pi a^3 \langle e' \rangle \langle I' \rangle$, where $a$ is the semi-major axis of Earth, 1~AU.  Under the assumption that $\langle \Delta v/v_k \rangle$ is relatively small, the mean eccentricity and inclination will be given by:
\begin{equation}
\langle e' \rangle \approx \sqrt{\frac{5}{3}} \left< \frac{\Delta v}{v_k} \right>,
\label{meaneprim}
\end{equation}
and
\begin{equation}
\langle I' \rangle \approx \sqrt{\frac{1}{3}} \left< \frac{\Delta v}{v_k} \right>.
\label{meanIprim}
\end{equation}
As this characteristic toroid, with cross-section centred at 1~AU, describes the mean of the orbital distribution we can expect that at any time roughly half of the debris will lie within the toroid.  We can thus expect that $n$ will be of order
\begin{equation}
\frac{N}{4 \pi^2 a^3 \langle e' \rangle \langle I' \rangle} \approx \frac{3}{4\sqrt{5} \pi^2 a^3} N \left< \frac{\Delta v}{v_k} \right>^{-2} \rm{particles~m^{-3}.}
\label{meandens}
\end{equation}

We can expect that the typical relative velocities at which debris encounters Earth will be similar to the velocity at which the debris was launched.  Thus we expect that $\langle v_{\rm{rel}} \rangle \approx \langle \Delta v \rangle $.

\begin{figure}
\includegraphics[width=85mm]{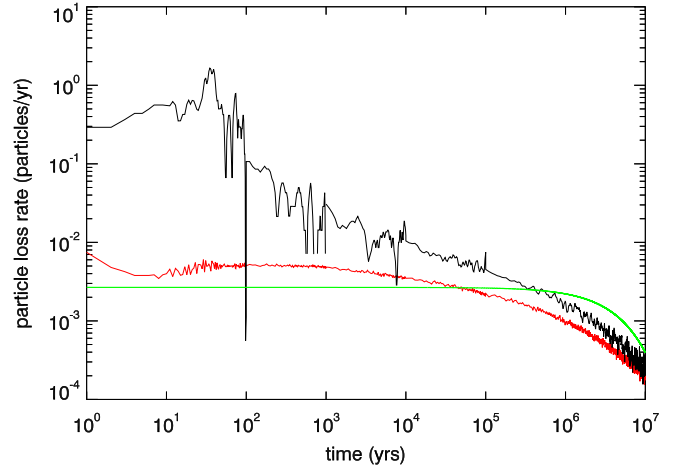}
\caption{Rate of loss of particles from the DE simulation (very few particles are ejected so this is essentially the rate of accretion).  In black is the actual (smoothed) rate of particle loss as calculated numerically from the simulation.  As the loss of individual particles is an inherently stochastic process this is somewhat noisy despite the smoothing applied.  We show in red the predicted accretion rate for Earth, $R_{\rm{acc}}$, with $n$ and $\langle \sigma v_{\rm{rel}} \rangle$ calculated at each time dump.}
\label{accrates}
\end{figure}

Checking against the DE simulation we find that these estimates do not quite match the real values, for reasons that are easily determined.  The density of particles is rather strongly peaked around Earth's orbit and thus the density very close to Earth's orbit will be somewhat higher than the mean density within our analytic toroid.  If we take a much smaller toroidal region around Earth's orbit with a cross-radius of 0.03~AU, or $\sim$3 Hill Radii (as a compromise between taking as tight a region as possible and avoiding low number statistics), we find that the density is around 3.5 times higher than estimated by Eq.~\ref{meandens}.  Similarly the mean relative velocity of particles within this same very narrow toroidal region is slightly lower than our estimate, at roughly $\frac{2}{3} \langle \Delta v \rangle$.

Using these corrected estimates we can provide an analytical estimate of the accretion rate as
\begin{equation}
R_{\rm{acc}} \approx \frac{7 R_{\oplus}^2}{4 \sqrt{5} \pi a^3} N \left( v_k \left< \frac{\Delta v}{v_k} \right> ^{-1} + \frac{9}{4}\frac{v_{\rm{esc}}^2}{v_k} \left< \frac{\Delta v}{v_k} \right> ^{-3} \right).
\label{analyticRacc}
\end{equation}
Plotting the solution to $\dot{N}=-R_{\rm{acc}}$ in Fig.~\ref{accrates} then produces the green curve.

Comparing the green analytic curve with the real accretion rate in Fig.~\ref{accrates} we see that at early times the true accretion rate is as much as two orders of magnitude higher than the analytic prediction as a result of the effects described in Section~\ref{peracc}.  At later times however, when the disk has become axisymmetric and the periodic effects are no longer significant, the analytic prediction of the accretion rate is much closer to the reality, as would be expected.  The green and black curves differ slightly at late times due to the necessity that the green curve is based only on the initial semi-major axis - eccentricity - inclination distribution of the debris, whereas in reality this does evolve slightly over time.

To confirm the origin of the difference between the analytic and real curves we can also calculate an `estimated axisymmetric accretion rate' at all times by calculating $n$ and $\langle \sigma v_{\rm{rel}} \rangle$ from the simulation at each timestep.  For this we use the 0.03~AU cross-radius toroid used to determine the correction to the analytic estimates above.  This results in the red curve in Fig.~\ref{accrates}, which is indeed close to the green curve at early times, confirming that the difference is due to the non-axisymmetric and periodic effects described in Section~\ref{peracc}.

\begin{figure*}
\includegraphics[width=85mm]{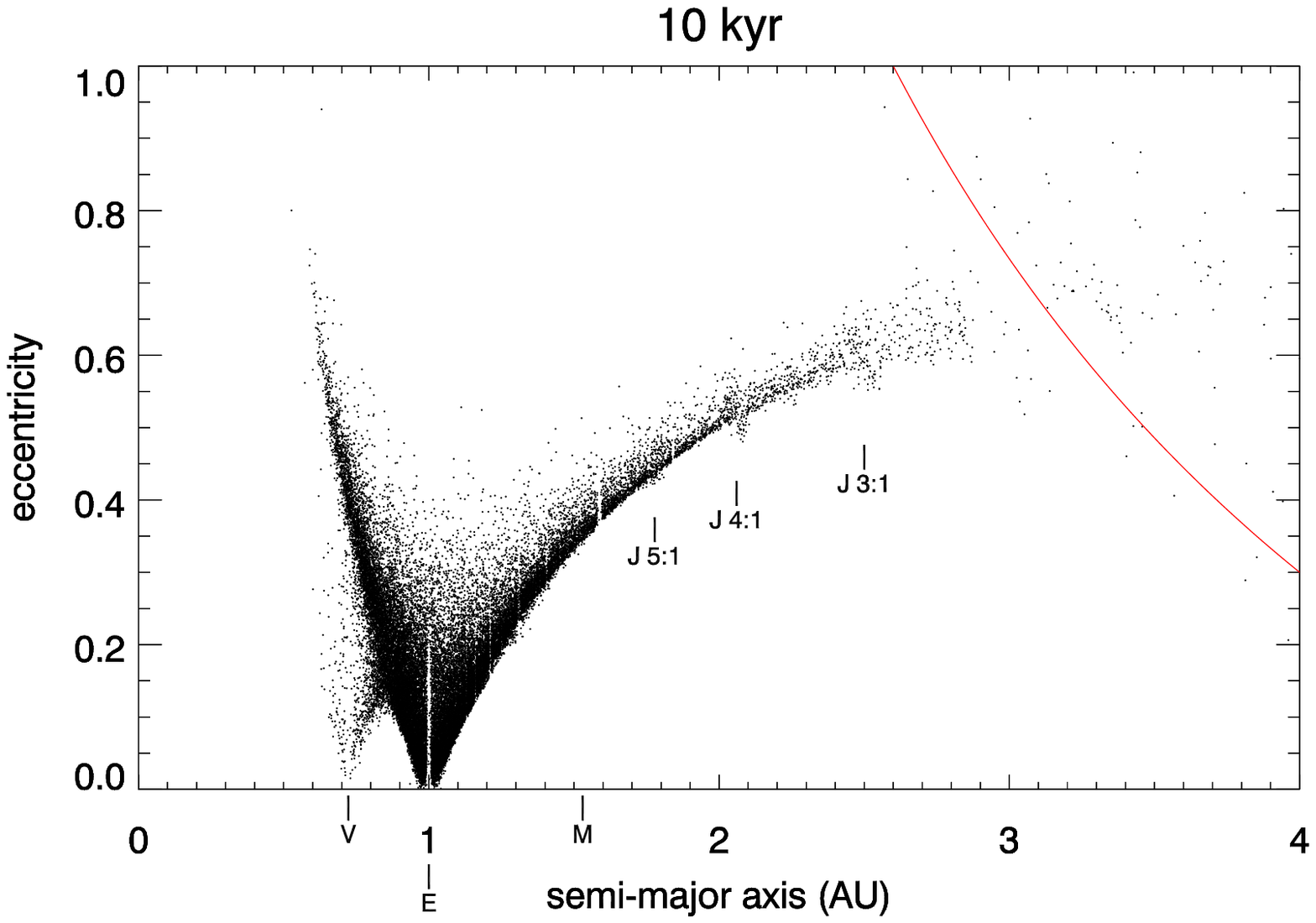}
\includegraphics[width=85mm]{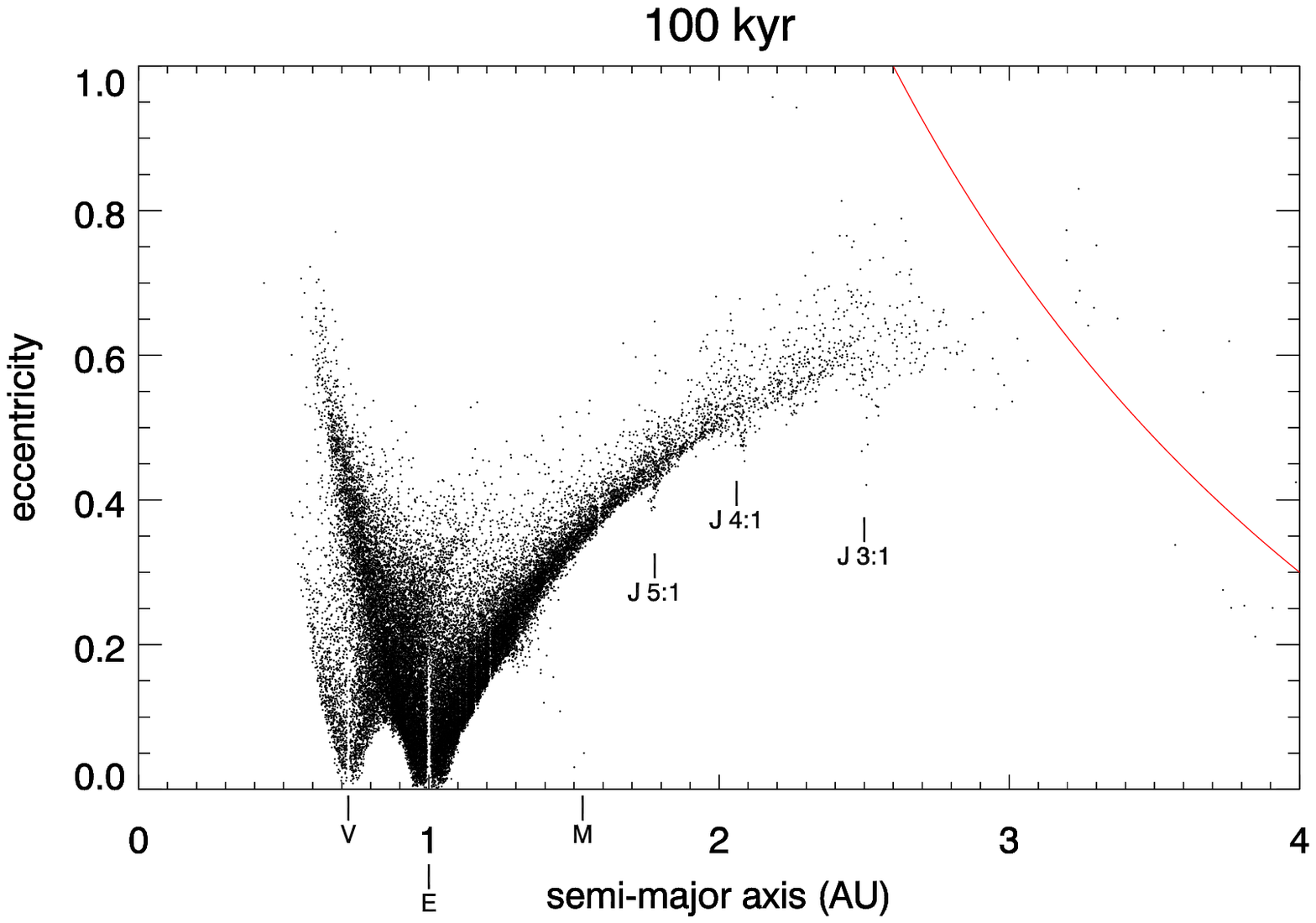}
\includegraphics[width=85mm]{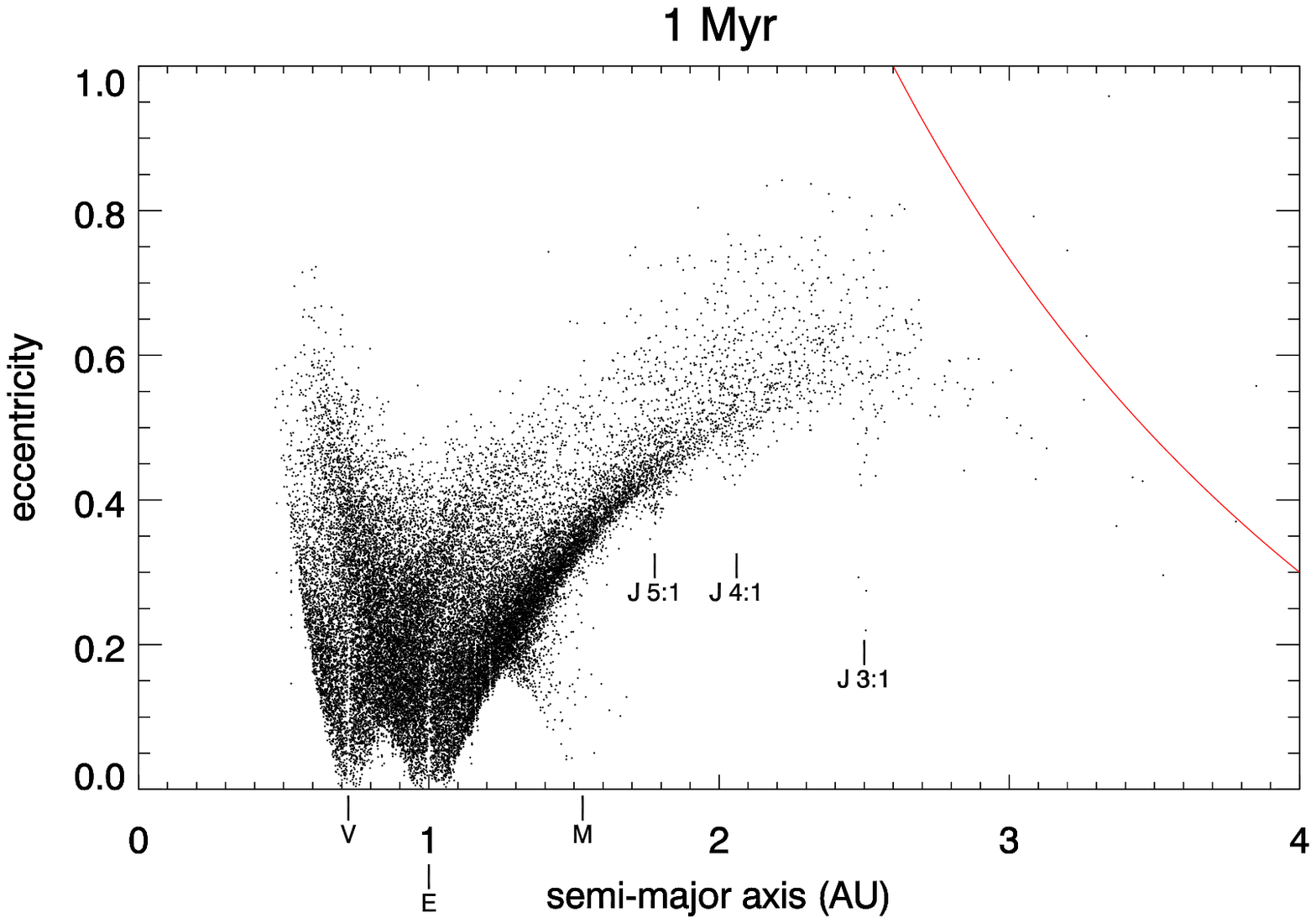}
\includegraphics[width=85mm]{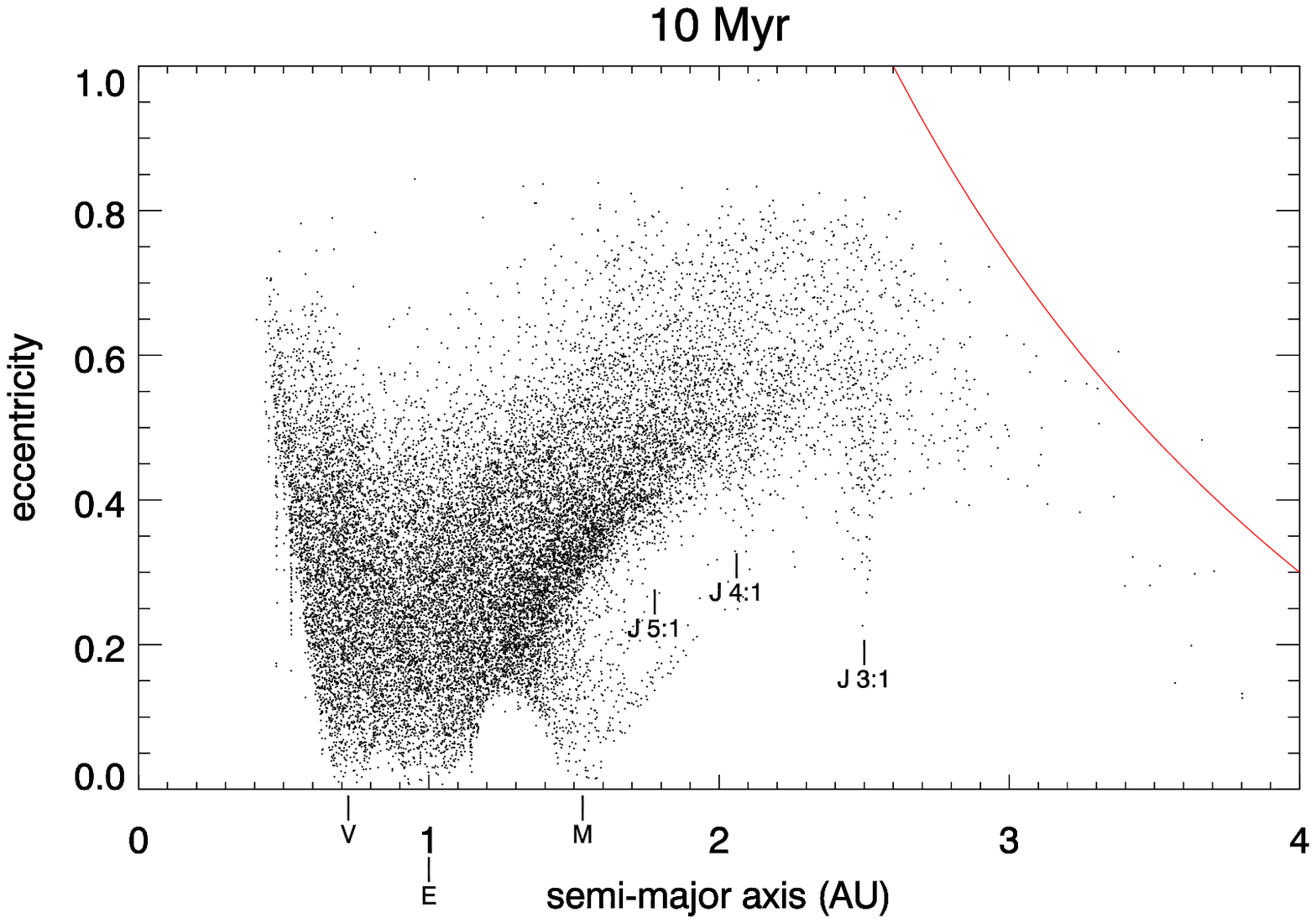}
\caption{Snapshots of the eccentricity -- semi-major axis distribution of the debris particles for the VEMJ simulation at 10~kyr, 100~kyr, 1~Myr and 10~Myr.  The locations of the 3:1, 4:1 and 5:1 resonances with Jupiter are marked, as are the positions of Venus, Earth and Mars. The red line indicates the locus of orbits for which the apocentre is located at 5.2 AU -- the orbital distance of Jupiter.  The build up of populations of particles associated with Venus and Mars, as well as the Jovian resonances, is apparent, as is the clearing of particles on Jupiter crossing orbits and the eccentricity pumping of particles with $a\ga2$ AU.}
\label{VEMJaedist}
\end{figure*}

\subsection{Adding in the other solar system planets}
\label{addAP}

We now consider the effects of adding in the other solar system planets.  The other planets which have noticeable influence on the evolution of the disk are, in descending order of magnitude, Venus, Jupiter and -- to a lesser extent -- Mars.  The effects of Mercury and planets beyond Jupiter are negligible.  The results of our final, highest resolution, simulation with 36000 particles and including Venus, Earth, Mars and Jupiter (which we refer to as the VEMJ simulation) are shown in Figs.~\ref{VEMJaedist} -- \ref{VEMJradhist}.  All of the additional planets are placed at their present semi-major axes but with zero mutual inclination and near circular orbits ($e\la10^{-3}$).  The initial true anomalies are randomised.  Such a set-up is beneficial in reducing the number of degrees of freedom in the system.  In addition the current mutual inclination and eccentricity state of the orbits of the Solar System planets are not necessarily representative of that at 50~Myr after the formation of the solar system, both since the planets undergo secular interactions over long timescales and models of the formation of the Solar System such as the Nice model (e.g. \citealt{tsiganis2005, brasser2009, morbidelli2009}) suggest that the orbits of the planets have changed since formation.  In addition the present mutual inclinations of the Solar System planets ($\la3^{\circ}$) are less than the typical inclinations of the debris particles (see Fig.~\ref{vrcomp}) and as such using the present mutual inclinations of the planets would not significantly change the mutual inclinations between the planets and the debris.  Similarly, although non-zero, the present eccentricities of the Solar System planets are not large enough to be likely to have significant effects.  The only planets for which the present non-zero eccentricity initially seems that it might have an effect are Mars and Mercury since their eccentricities of 0.1 and 0.2 bring them 10 and 20 per cent closer to 1~AU, into denser parts of the debris disk.  However the time they spend closer to 1~AU is largely offset by the time they spend further from 1~AU; for example, their mean orbital distances, $\langle a \rangle = a(1+\frac{1}{2}e^2)$, differ from their semi-major axes by only 0.5 and 2 per cent respectively.  

\subsubsection{Mercury}
\label{addMercury}

The small semi-major axis of Mercury means that only particles with high eccentricities can approach the planet closely enough to interact with it.  This significantly reduces the proportion of the population that can interact with the planet.  Further the low mass of Mercury, and the high relative velocity that such eccentric particles will encounter the planet with, mean that the cross-section for interactions between Mercury and those disk particles which have the potential to interact with it is rather small.  As a result the influence of Mercury on the evolution of the disk is very small.  This is confirmed by shorter, lower resolution simulations including all of the solar planets. In the VEMJ simulation we thus do not include Mercury to ease the computational burden.

\subsubsection{Jupiter and the outer planets}
\label{addJupiter}

In contrast with the effect of Earth described in Section \ref{DEdyn}, the effect of Jupiter on those particles which interact with it is predominantly through scattering, which typically leads to the ejection of material from the solar system.  As is the case with Mercury, for material ejected from the Moon-forming impact to interact closely with Jupiter it must have been put onto an orbit with a high eccentricity and will thus encounter Jupiter with a substantial relative velocity.  Combined with the large mass of Jupiter this means that an object undergoing an encounter with Jupiter is likely to receive a large kick to its velocity, which frequently leads to the ejection of the particle.  Unlike Mercury, the large mass of Jupiter leads to a very large gravitational focussing effect enlarging the interaction cross-section.  Thus, although the fraction of the population that can undergo interactions with Jupiter is small, the likelihood of particles within that fraction undergoing an interaction is high.  This means that most particles on Jupiter-crossing orbits will undergo a strong encounter with Jupiter, and most likely be ejected from the solar system, on relatively short timescales.  As well as ejecting objects from the solar system entirely, scattering with Jupiter is also important for throwing material into the innermost regions of the solar system.  In simulations lacking Jupiter no particles are observed to impact the Sun for example (0/16000), whereas a small number (0.08\%, 27/36000 particles) do when it is included.

This ejection of particles on Jupiter-crossing orbits results in the disk being truncated at a radius just interior to the orbit of Jupiter, which is evidenced in Fig.~\ref{VEMJaedist} by the lack of particles occupying the region above the red line where orbits will intersect that of Jupiter.  This also has consequences for interactions with the other outer planets, since in order for a particle to undergo a close encounter with one of the other outer planets it will also have had to cross the orbit of Jupiter.  Being substantially more massive than the other outer planets, any object on an orbit that could potentially interact with one of the other outer planets is significantly more likely to interact with Jupiter and be ejected.  The other outer planets are thus prevented from playing a significant part in the evolution of the disk by the presence of Jupiter.  Although today Saturn plays a role in controlling the structure of the asteroid belt (e.g. \citealt*{murray1998, nesvorny1998}) through long range gravitational effects rather than direct close encounters the timescales we are investigating here are too short for these to become important.   As with the very small influence of Mercury this is confirmed by simulations containing all of the solar system planets, and so the planets beyond Jupiter are not included in the VEMJ simulation to reduce the computational burden.

In addition to the ejection from the solar system of objects on Jupiter-crossing orbits the large mass of Jupiter means that the influence of mean-motion resonances with Jupiter is noticeable, for example, in Fig.~\ref{VEMJaedist}, a population of particles occupying the 3:1 mean-motion resonance with Jupiter is distinguishable.  Earlier in the evolution, distinct populations of particles occupying the 4:1 and 5:1 resonances at 2.1 AU and 1.8 AU respectively are also clearly visible.  The Jovian 2:1 resonance is not visible since particles with a 2:1 period ratio with Jupiter that are on Earth-crossing orbits would also have to be on Jupiter-crossing orbits and so would be rapidly ejected.  Comparison of Figs.~\ref{VEMJaedist} and \ref{DEaedist} also shows weaker scattering up to higher eccentricities of particles on orbits that pass relatively close to, but do not cross, the orbit of Jupiter (those with $a \ga 2$ AU).  Over the 10~Myr integration time of the VEMJ simulation 2804 out of the 36000 initial debris particles are ejected, $\sim$8\%, almost entirely due to the influence of Jupiter.  As a result of the strong scattering Jupiter induces, the number of particles actually accreted onto Jupiter is rather low (0.3\%, 120/36000 particles).

\begin{figure}
\begin{center}
\includegraphics[width=60mm]{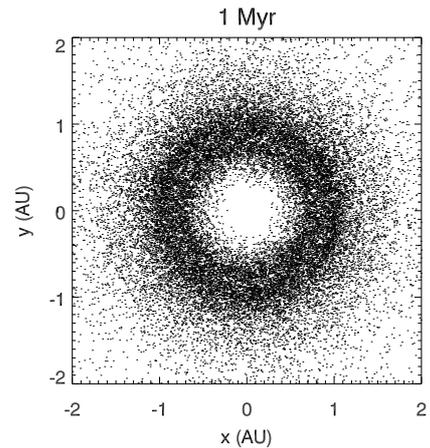}
\end{center}
\caption{Face-on view of the VEMJ simulation disk at 1 Myr.  The broadening of the region of high density into a band sitting across the orbits of Earth and Venus is readily apparent.}
\label{VEMJ1Myrxy}
\end{figure}

\subsubsection{Venus}
\label{addVenus}

As mentioned above Venus has a strong influence on the disk, second in magnitude only to that of Earth.  The strength of this influence is a result of the comparable mass of Venus to that of Earth and its relatively nearby orbital position.  Its nearby orbit means that Venus can interact with particles on orbits with relatively low eccentricities, which will thus have lower relative velocities to the planet during close approaches, while its mass means that its interaction cross-section is significant.

\begin{figure}
\includegraphics[width=85mm]{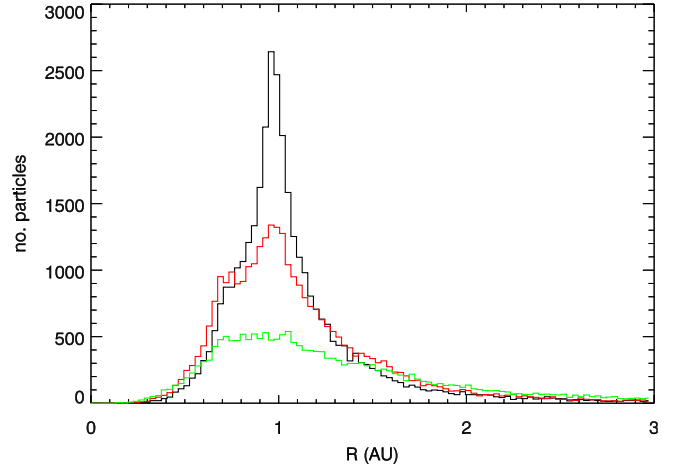}
\caption{Evolution of the radial density profile of the disk for the VEMJ simulation.  Snapshots shown at 100 kyrs (black), 1 Myr (red) and 10 Myr (green).  The density is strongly peaked at 1~AU at early times but this decreases with the peak also spreading inwards towards the orbit of Venus at later times as particles are re-accreted by the Earth and interact with the other planets.  A slight increase in the number of particles beyond around 1.4~AU due to the influence of Mars is also evident.  Compare with Fig.~\ref{DEradhist}.}
\label{VEMJradhist}
\end{figure}

Interactions between Venus and the debris particles result in the strong peak in density at 1 AU being spread inwards toward the orbit of Venus.  By 1 Myr there is a distinct band of higher density in the disk that spans the region between the orbits of Venus and Earth, as seen in Figs.~\ref{VEMJ1Myrxy} and \ref{VEMJradhist}.  By 10 Myr interactions with Venus have completely smoothed the distribution between the orbits of Earth and Venus with Figs.~\ref{VEMJaedist} and \ref{VEMJradhist} clearly showing a large population of particles that have interacted with Venus and become associated with it.  The scattering of debris particles by the planets is governed by the same equations as those that determine the initial distribution of the debris.  This is clearest in the distinctive `V' shape produced in the eccentricity -- semi-major axis distribution which allows us to distinguish particles associated with Venus in Fig.~\ref{VEMJaedist} as those which lie in the `V' centred at the orbit of Venus.

As well as significantly altering the structure of the disk Venus also accretes large numbers of particles.  The amount of material accreted onto Venus by the end of the VEMJ simulation run at 10 Myr (6016 particles, 17\% of the total) is comparable to the amount of material re-accreted by Earth.  At 7142 out of 36000 (20\%) the number of particles re-accreted onto Earth is in turn significantly lower than the DE simulation.  The finding that Venus accretes a comparable amount of material from the disk to Earth can be compared with the results of studies of the evolution of ejecta from impacts onto the terrestrial planets such as \citet{gladman1996}, who also found that after a few hundred thousand years the rate of collision of material launched from the vicinity of Earth (specifically from the Moon in their study) onto Venus was the same as that onto Earth.

\subsubsection{Mars}
\label{addMars}

As a terrestrial planet the interaction of Mars with the disk is similar to that of Venus and Earth. The greater orbital separation between Earth and Mars increases the necessary eccentricity for particles to be able to interact with Mars and this, as well as Mars' lower mass, reduces the interaction cross-section of Mars.  As a result the influence of Mars on the disk is rather smaller than that of Venus or Earth.  Nonetheless the influence of Mars still produces a noticeable impact on the disk unlike Mercury, and a population of particles associated with Mars is clearly distinguishable in Fig.~\ref{VEMJaedist}, albeit that this population is significantly smaller than that associated with Venus or Earth.  The number of particles actually accreted onto Mars is also much lower (0.3\%, 102 particles) than the number accreted onto Venus or Earth.  The low accretion rate is likely partly due to a longer timescale for material to become strongly associated with Mars, for example the region around Mars at 10 Myr is rather similar to that around Venus at 10 kyrs.  While the proximity of Jupiter and strong Jovian resonances might be expected to have a noticeable effect when considering disk-Mars interactions, this is not observed.  In simulations lacking Jupiter but retaining Mars there are no significant changes in the low eccentricity ($e\la0.2$) distribution near Mars.

\subsubsection{Cumulative effects of the other solar system planets}
\label{addcum}

As a result of the increased stirring of the disk with additional planets, despite the significant accretion of material onto Venus and increase in number of particles ejected, the total number of particles lost over 10 Myr is decreased by around 20 percent when Venus, Mars and Jupiter are added.  Thus the lifetime of the disk particles for loss to dynamical interactions is actually increased by the addition of the other solar system planets.  This is because re-accretion onto Earth is decreased dramatically (from 58\% of particles to 20\% of particles) by the increased stirring and although, in the case of Jupiter and Venus at least, the other planets can remove substantial numbers of particles from the disk it is insufficient to compensate for this reduction.  Fitting the same function, $N = N_0 \exp(-\sqrt{t/\tau})$, to the number of particles remaining in the disk, $\tau$ is increased to 29~Myr with the other solar system planets present.

\begin{figure}
\includegraphics[width=\columnwidth]{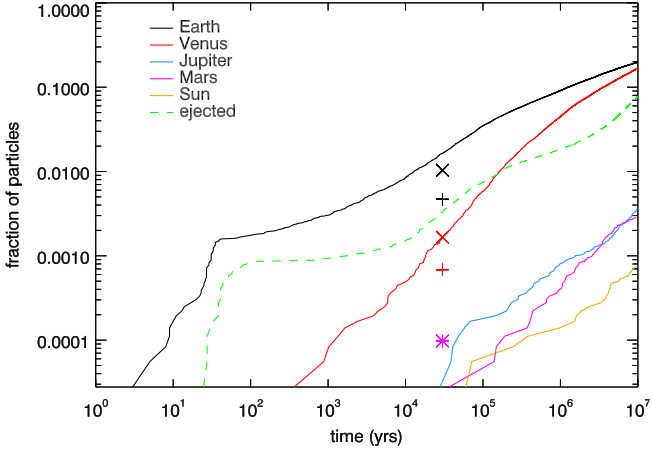}
\caption{Fraction of particles lost from the VEMJ simulation over time to accretion onto different bodies and ejection from the Solar System.  For comparison we plot the results of \citet{reyes2011} for their cases B and C (as described in the text), as crosses and pluses respectively in the same colour as the corresponding lines.}
\label{Nacc}
\end{figure}

In Fig.~\ref{Nacc} we show the numbers of particles either accreted onto the different planets or ejected from the solar system.  The very steep initial phase of accretion onto Earth is due to the initial clearing of the commensurability gaps (see Section \ref{peracc} and Fig.~\ref{accrates}) while the very steep initial phase of ejection is due to particles placed on initial orbits with $e>1$ leaving the simulation.

\begin{figure}
 \includegraphics[width=\columnwidth]{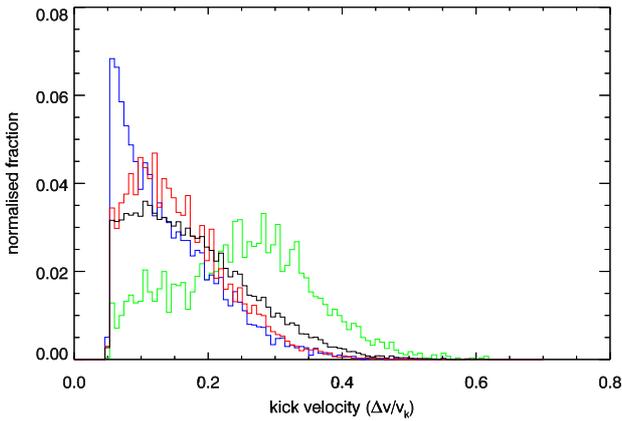}
 \caption{Distributions of initial kicks received by particles that are subsequently re-accreted onto Earth (blue), accreted by Venus (red) or ejected from the Solar System (green) over the 10~Myr of the VEMJ simulation as compared to the initial distribution for all particles (thick black).  The curves are normalised such that the area under each is equal.}
 \label{dvdesthist}
\end{figure}

\citet{reyes2011} conduct a series of shorter, smaller, simulations (30~kyrs, 10242 particles) with fixed launch velocities.  At their launch altitude of 100~km the mean initial velocity imparted to our particles would correspond to a mean launch velocity of 12.2~km~s$^{-1}$, part way between their cases B and C.  We show their results for these two cases in Fig.~\ref{Nacc} as crosses and pluses respectively.  Their results compare well to ours, with minor deviations due to small number statistics in their results (their results for Mars are based on a single particle for example), and the distribution of velocities we use versus their fixed velocities.  The lower velocities present in our distribution lead to increased accretion onto Earth, while the high velocity tail increases interaction with Jupiter and ejection.  This is illustrated by Fig.~\ref{dvdesthist} which shows the distributions of initial kick velocities received by particles that are lost to each of the three main sinks (accretion onto Earth or Venus and ejection) and in which it is clearly visible that particles re-accreted onto Earth are preferentially those that received lower initial kicks, while those ejected are preferentially those that received larger initial kicks.  In addition to the preference for having received larger initial kicks, particles which are ejected also show a strong preference for having been kicked in the direction of orbital motion, since this increases the semi-major axis, as was also found by \citet{reyes2011}.

From the differences between the DE and VEMJ simulations it is clear that, although the most important factor in the dynamical evolution of debris from a Moon-forming type impact is its parent planet, the wider planetary environment in which that parent exists is also very important.  The largest such influence is the presence of other nearby planets of comparable mass to the parent.  The presence of other planets in the system also causes the disk to become axisymmetric faster and decreases the length of the phase in which the periodic scattering and accretion described in Section \ref{peracc} is dominant.

With the addition of eccentricity pumping by Jupiter and trapping in Jovian resonances the eccentricity -- semi-major axis distributions of Fig.~\ref{VEMJaedist} are well described in terms of overlapping `V's centred at the orbital distances of Venus, Earth and Mars.  These `V' shapes, associated with orbits that have an apocentre or pericentre at the location of the planet, also determine the fraction of the initial debris distribution with which a planet can interact.

\section{Collisional evolution of the debris}
\label{collisions}
In addition to dynamical evolution due to the external influence of the planets, the debris from a Moon-forming or other giant impact will also evolve through self-collision.  To determine the collisional evolution it is necessary to know the size distribution of the debris, whereas the dynamical evolution is blind to the size distribution (provided the largest debris particles are small enough for gravitational focussing within the debris distribution to be neglected).

\subsection{The size distribution}
\label{sizedist}
As mentioned in Section \ref{initconds} the majority of escaping particles in the \textsc{sph} simulation that informs our initial conditions occurred as part of small bound associations (fragments).  The largest such fragments consisted of 6 particles and had a mass of $2.6 \times 10^{21}$~kg, translating to a diameter of 570~km assuming a density of 3300~kg~m$^{-3}$, similar to that of the Moon.  However with only 6 particles there is considerable uncertainty in this size and notably there is a risk of spurious associations when dealing with such small groups of \textsc{sph} particles.  Nonetheless this does at least set a rough upper limit on the size of the largest objects in the distribution as $\sim$600km.

In an impact simulation similar to that on which our present analysis is based, \citet{canup2008} (their Fig.~9) note that the largest fragment in the proto-lunar disk 27 hours after impact, at which time we would expect negligible re-accumulation to have occurred, has a mass of 0.16$M_{\rm{L}}$.  This corresponds to a diameter of $\sim$1800km at lunar density.  Although we do not find any fragments this large in the ejected material it is expected that the proto-lunar disk would contain larger objects than the ejected material since the largest fragments should generally be imparted slightly lower initial velocities (\citealt{leinhardt2011}).

Furthermore we can use the method of \citet{wyatt2002} to estimate the size of the second largest remnant given the total mass of debris, both unbound and bound in the proto-Lunar disk.  This method gives the size of the second largest remnant, $D_{\rm{slr}}$, (the largest body in the debris distribution, with Earth as the largest remnant) as
\begin{equation}
D_{\rm{slr}} = \left[\frac{3 M_{\rm{tot}}}{\pi\rho}\frac{4-\alpha}{\alpha-1}\left(1-\frac{M_{\rm{lr}}}{M_{\rm{tot}}}\right)\right]^{\frac{1}{3}}
\end{equation}
where $M_{\rm{tot}}$ is the total colliding mass, $M_{\rm{lr}}$ is the mass of the largest remnant (Earth), $\rho$ is the density of the debris material (assumed to be Lunar) and $\alpha$ is the power law index in the size distribution $n(D)dD \propto D^{-\alpha}dD$.  Around 3.5\% of the total colliding mass is converted into debris, either in the proto-Lunar disk or in the unbound debris, and assuming $\alpha$=7/2, as for a self-similar collisional cascade (this will be discussed in more detail in Section~\ref{sizedistshape}), suggests a second largest object size, $D_{\rm{slr}}\sim2200$~km.  This is very similar to the size of the large object in the proto-Lunar disk of \citet{canup2008}, suggesting this is not a spurious occurence.  If we assume that the largest debris objects remain in the proto-Lunar disk and integrate down from a largest object of $\sim$2000~km until we have accounted for the mass of the proto-Lunar disk, we then find that the largest objects in the unbound debris should be $\sim$500~km.  While certainly not conclusive evidence, this supports the potential existence of large fragments with diameters of hundreds of kilometres in the unbound debris.

\subsubsection{Vapour}
\label{vapour}
In addition to these large fragments some of the material thrown out in the impact will have been subjected to sufficient pressure and temperature to have been vapourised.  The fraction of material that is vapourised is estimated by \citet{canup2008} as $\sim$10-30 percent by mass.  This vapourised material will cool and recondense along pathways determined by the exact conditions to which it is subjected during the impact, but the largest resulting droplets are expected to have sizes ranging from mm to cm scales (e.g. \citealt{melosh1991}, \citealt{anic2006}, \citealt{benz2007}).  We will thus have a two component distribution comprised of small material that recondensed from vapour and larger material that was subjected to less violent conditions.  We will, hereafter, refer to material arising from the condensation of vapour as the `vapour distribution' or `vapour material' and to material that was not vapourised as the `boulder distribution' or `boulder material'.

\subsubsection{The shape of the size distribution}
\label{sizedistshape}
While from the \textsc{sph} simulation we have some estimates of the size of the largest fragments in the boulder distribution, albeit with uncertainties as described above, it does not really give us any information about the shape of the size distribution.  Even a single \textsc{sph} particle would translate into an object $\sim$300 km in diameter.  In order to proceed we make the assumption that the size distribution of boulder material obeys a power law with a slope equal to that of a steady-state self-similar collisional cascade, so that $n(D)dD$, the number of objects between size $D$ and $D+dD$ is $\propto D^{-7/2}dD$ (\citealt*{tanaka1996}).  As a result, rather than each particle in the dynamical simulations representing an individual debris fragment, it represents an ensemble of fragments conforming to this size distribution.

It is reasonable to expect that the small objects in the cascade, which have short collisional lifetimes, will rapidly reach a steady-state so that this size distribution will be a relatively accurate reflection of the reality (\citealt*{wyatt2011}).  For larger, longer lived objects this is less certain, but we have no evidence for a different distribution and this is the simplest assumption.  This simple steady-state distribution also has the advantage that it will remain the same as the material evolves both collisionally and dynamically (since the dynamics is independent of the fragment size).  For a size distribution with this slope most of the mass is contained in the largest objects and thus the rate at which mass is lost from the distribution by collisions is determined by the lifetimes of the largest objects.  These arguments apply equally to both the boulder and vapour distributions, the main difference being the size of the largest objects in the respective distributions.

Note that in the more realistic situation that planetessimal strength ($Q_D^*$) is a function of size (see Section~\ref{collevol}) the steady-state equilibrium size distribution will have a slope that may differ from the $7/2$ assumed here (e.g. \citealt{wyatt2011}).  However the simplicity of the assumption of a single power law slope, given other uncertainties about the size distribution, in particular $D_{\rm{max}}$, make this a reasonable approximation.

\subsection{Evolving the collisional cascade}
\label{collevol}
Our dynamical simulations provide us with the distribution of orbital parameters of the debris at each time-dump, which gives us the basis for calculating the probabilities for collision between debris particles.  To determine the probability of a catastrophic collision for an object of size $D$, i.e. the probability of a collision in which the largest remnant will have less than half of the mass of the initial object, we require the specific incident energy necessary to disperse this mass (dispersal threshold), $Q_D^*$.  The dispersal threshold energy allows us to calculate $D_{cc}$, the size of the smallest object capable of participating in a catastrophic collision with an object of size $D$ for the typical collision velocities ($v_c$) present, and thus the fraction of collisions that will be catastrophic.  The size we expect for the largest objects in the boulder distribution places them firmly in the gravity regime where it is primarily the self-gravity of the object, rather than its material strength, that determines its resistance to dispersal.  On the other hand the vapour distribution lies entirely in the strength regime.

Utilising the velocity-dependent dispersal threshold of \citet{stewart2009} we obtain,
\begin{equation}
\label{QD*}
Q_D^* = q_s \rho^{-0.109}(D/2)^{-0.327}v_c^{0.8} + q_g \rho^{0.4}(D/2)^{1.2}v_c^{0.8},
\end{equation}
where $\rho$ is the density of the debris material (assumed to be Lunar at 3300 kg~m$^{-3}$), $v_c$ is the collision velocity, $D$ is the diameter of the object.  \citet{stewart2009} find that for weak aggregates, $\mu=0.4$, $\phi=7$, and in SI (mks) units $q_s=0.937$ and $q_g=6.31\times10^{-6}$.  Note that \citet{stewart2009} define the dispersal threshold in terms of $Q_{RD}^*$, the specific incident energy necessary to disperse half of the \emph{combined mass of the target and projectile}, rather than $Q_D^*$.  For our purposes however the difference between $Q_{RD}^*$ and $Q_D^*$ is relatively small as the typical velocities at which collisions occur in the debris disk are, even for the largest objects, substantially larger than the escape velocities of the objects and thus the minimum mass of a catastrophically destructive projectile is much less than that of the target.

Armed with this we can then interweave collisional and dynamical evolution by calculating the mean collision velocities and the collisional lifetime of the largest objects at each time-dump of the dynamical simulation.  The dynamical mass loss is tracked simply by the number of particles remaining in the simulation, $N$, compared to the initial number of particles, $N_0$.  As the dynamical evolution is independent of the size of individual debris fragments each dynamical particle simply represents an equal fraction of the remaining disk mass.  Initially each dynamical particle will thus represent a mass $m_0=M_0/N_0$, where $M_0$ is the total initial mass of debris.  Then, since mass is lost from the cascade through the destruction of the largest objects, this allows us to determine the mass lost through collisional processes over the period between successive data dumps of the dynamical simulations.  The mass per dynamical particle at the beginning of the $i$th time period, $m_i$, is determined by
\begin{equation}
\label{collcalc}
m_i=m_{i-1}\frac{1}{1+(t_i-t_{i-1})/t^c_{i-1}}
\end{equation}
where $t_i$ is the time at the beginning of the $i$th time period and $t^c_{i-1}$ is the collisional lifetime of the largest object (size $D_{\rm{max}}$) at the beginning of the $(i-1)$th time period.  

The collisional lifetime of the largest object, $t^c$, (equivalently the catastrophic collision timescale for the largest object) is given by
\begin{equation}
\label{collifetime}
t^c=\frac{1}{P_c \sigma_{cc}(D_{\rm{max}})}.
\end{equation}
Here $P_c$ is the intrinsic collision probability (m$^{-2}$ s$^{-1}$), determined numerically from the distribution of orbital parameters of the debris from the dynamical simulation using the method of \citet{wyatt2010} (see also \citealt{bottke1994}), and $\sigma_{cc}(D_{\rm{max}})$ is the cross-section for catastrophic collision for an object of size $D_{\rm{max}}$.  The catastrophic collision cross-section, $\sigma_{cc}(D_{\rm{max}})$, is 
\begin{equation}
\label{sigmacc}
\sigma_{cc}=\int_{D_{cc}(D_{\rm{max}})}^{D_{\rm{max}}} n(D) \left( \frac{D_{\rm{max}}+D}{2} \right) ^2 dD,
\end{equation}
where $D_{cc}(D_{\rm{max}})$ is the size of the smallest object capable of catastrophically colliding with an object of size $D_{\rm{max}}$ (see e.g. \citealt{wyatt2002}).  We neglect gravitational focussing within the debris distribution, since the typical collision velocities of $\ga$5~km~s$^{-1}$ are much larger than the escape velocity of even 1000~km objects.

\subsubsection{Assumptions of the collisional model}
\label{collasump}

These calculations assume that the disk is axisymmetric.  This is true at later times but in the early phases of the evolution this assumption breaks down and the collisional evolution will not be correctly determined by this prescription.  Figs.~\ref{DEphasenc} and \ref{accrates} show that it is during the first few kyrs that the effect of the non-axisymmetry is strongest after which the effect fades.  Where the timescales for collisional evolution are longer than this initial period the non-axisymmetry should not significantly influence the evolution.  This is a similar timescale to the initial collisional lifetime for objects at the transition from the strength regime to the gravity regime ($D_{\rm{max}}\sim 0.1-1$~km), so we expect that the early asymmetric period will be important for distributions where $D_{\rm{max}}$ lies in the strength regime, while distributions that extend into the gravity regime will be largely unaffected.

In order to extend our collisional model to times later than 10~Myr (necessary to follow distributions with large $D_{\rm{max}}$ to the time at which they become undetectable) we make the assumption that the orbital distribution at later times can be approximated by that at 10~Myr.  Though the orbital distribution is likely to continue to evolve past 10~Myr, the rate of evolution is likely to slow as the particles with the shortest dynamical lifetimes are naturally those that are lost first.  Material will also continue to be lost from the disk through dynamical effects (primarily accretion) after 10~Myr.  In order to account for this we perform a fit to the particle loss rate from 1-10~Myr, removing the first Myr to remove the effects of the early non-axisymmetry.

\subsubsection{Luminosity of the disk}
\label{obsdisk}

From \citet{wyatt2008} the fractional luminosity (ratio of disk to stellar luminosity) of a collisional cascade with a size distribution $n(D)\propto D^{-7/2}$ is related to the total mass of the cascade by:
\begin{equation}
\label{fraclumeq}
f=0.37 \left(\frac{M_{\rm{tot}}}{M_{\oplus}}\right) \left(\frac{r}{AU}\right)^{-2}\left(\frac{D_{\rm{bl}}}{\mu \rm{m}}\right)^{-\frac{1}{2}}\left(\frac{D_{\rm{max}}}{\rm{km}}\right)^{-\frac{1}{2}}
\end{equation}
where $D_{\rm{bl}}$ is the minimum size of dust grains in the distribution, which is expected to be set by radiation pressure at $D_{\rm{bl}}=0.65(L_*/L_{\odot})(M_{\odot}/M_*)(3300 \rm{kg}\hspace{4pt}\rm{m}^{-3}/\rho)$~$\mu$m.  Thus for our assumptions $D_{\rm{bl}}=0.65 \mu$m.  The typical radius of the disk is $r$, which here is $\sim$1 AU, and being peaked at this distance the dust will be hot ($\sim$300 K) with emission thus peaking at $\sim$10~$\mu$m.  The detectability of the disk emission at a particular wavelength, $F_{\lambda}^{\rm{disk}}$, is determined by how bright it is relative to the stellar flux, $F_{\lambda}^*$.  At a wavelength $\lambda$, the disk results in an excess of emission above that expected from the stellar photosphere and so we define the fractional excess, $R_{\lambda}=F_{\lambda}^{\rm{disk}}/F_{\lambda}^*$, or equivalently $R_{\lambda}=(F_{\lambda}^{\rm{tot}}-F_{\lambda}^*)/F_{\lambda}^*$.  In the context of past surveys for debris disks, such as \citet{trilling2008} or \citet{carpenter2009}, this would result in the strongest detection being in the 24~$\mu$m band, for which $R_{24}\approx 3200f$.  For calibration limited surveys such as these the limit of detectability is typically $R_{24}=0.1$.

\subsection{Changing the top of the size distribution}
\label{changesizedist}

As discussed in Section~\ref{sizedist}, the size of the largest object in the distribution is subject to some uncertainty.  As such in this Section we apply the scheme described in Section~\ref{collevol} to the VEMJ simulation to investigate the variation of the collisional evolution of the distribution for a wide range of values of $D_{\rm{max}}$.

\subsubsection{Varying $D_{\rm{max}}$}
\label{varyDmax}

\begin{figure}
\includegraphics[width=85mm]{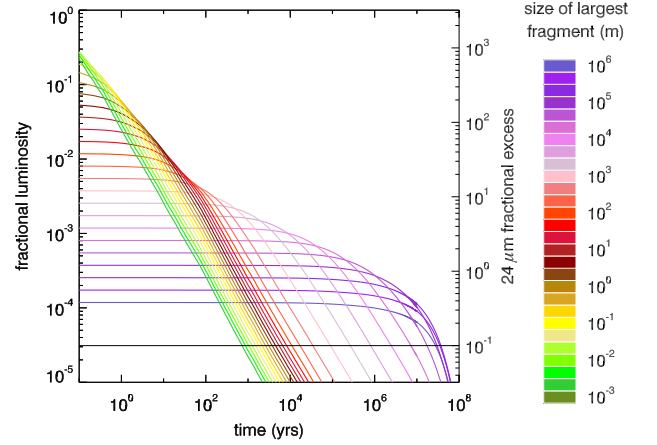}
\caption{Evolution of the fractional luminosity and 24~$\mu$m excess of the disk for different values of $D_{\rm{max}}$ in a single continuous distribution.  The lines are coloured by the value of $D_{\rm{max}}$ as shown in the colour bar at right.  In black we show a typical survey detection limit of $R_{24}=0.1$ (e.g. \citealt{carpenter2009, trilling2008}).}
\label{femul}
\end{figure}

Let us consider the effects of varying $D_{\rm{max}}$ for a single continuous distribution of initial mass 1.3~$M_{\rm{L}}$ (0.016~$M_{\oplus}$).  In Fig.~\ref{femul} we explore the entire range of $D_{\rm{max}}$ from millimetres up to thousands of kilometres, spanning the range of sizes expected for both the boulder and vapour distributions.

If we decrease $D_{\rm{max}}$ then, since the initial mass of the cascade is constant and we do not change $D_{\rm{bl}}$, Eq.~\ref{fraclumeq} states that the initial brightness of the disk must increase.  Physically we can see this must be the case, since we shift the entire mass of the cascade down to smaller objects and thus a greater proportion of the mass is in the observable small dust grains, meaning the disk must be brighter.  We can clearly see this in Fig.~\ref{femul}, and note also that the upper envelope of this plot gives the maximum possible brightness for a distribution containing initially 1.3 $M_{\rm{L}}$ of material and an $n(D)\propto D^{-7/2}$ shape.  Eq.~\ref{sigmacc} tells us that as we increase $D_{\rm{max}}$, $\sigma_{cc}$ must decrease since increasing $D_{\rm{max}}$ increases the proportion of the total mass contained in a single object of size $D_{\rm{max}}$ and the number of objects in the range $D_{cc}$ to $D_{\rm{max}}$ falls.  Thus as $D_{\rm{max}}$ is increased $t^c$ increases.  Again we see this in Fig.~\ref{femul} since distributions with a larger $D_{\rm{max}}$, while initially less bright, decay more slowly.

There are two kinks in the shape of the upper envelope of Fig.~\ref{femul}.  The first, at around 50~yrs (and maximum sizes of a few hundred metres) is due to the largest objects transitioning from the strength regime to the gravity regime.  The second kink, at around 10~Myr (and maximum sizes of around a hundred kilometres) is a result of dynamical evolution becoming the dominant process in controlling mass loss from the disk, rather than collisional grinding.

\begin{figure}
\includegraphics[width=85mm]{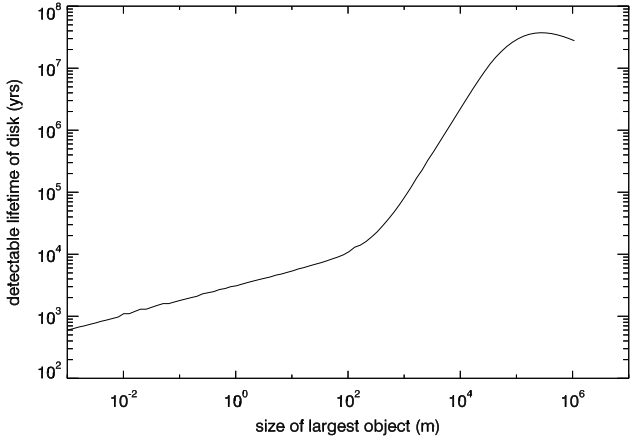}
\caption{Detectable lifetime of the disk as a function of the size of the largest object for the VEMJ simulation.}
\label{detecttime}
\end{figure}

The detectable lifetime of the disk, defined here as the length of time for the 24~$\mu$m excess to decay to a level of less than 0.1, is shown in Fig.~\ref{detecttime}.  As described above there is in general a decrease in the detectable lifetime of the disk as $D_{\rm{max}}$ is decreased, and there is again a kink at a $D_{\rm{max}}$ of several hundred metres due to the transition from the strength to the gravity regime. The curve also begins to turn over at several tens of kilometres, above which dynamical evolution dominates.  Dynamical evolution sets a maximum detectable lifetime for the debris of around 37~Myr with this peak occurring at 300~km.  For distributions with $D_{\rm{max}}$ larger than around 300~km the detectable lifetime slowly decreases as a result of the net lifetime of the largest objects now being constant and independent of collisional effects (determined only by dynamics), while the surface area of the distribution for the same mass decreases with increasing $D_{\rm{max}}$.

Note that at small values of $D_{\rm{max}}$, particularly those relevant to the vapour distribution, there are a number of issues that come into play which thus far we have not mentioned.  The structure of the disk during the non-axisymmetric phase (see Section \ref{peracc}) leads to a substantially higher re-accretion rate onto Earth at early times, and it can be expected that the asymmetry embodied in the concepts of the collision point and anti-collision line will similarly lead to higher collision rates than predicted in our axisymmetric calculations.  In addition, at values of $D_{\rm{max}}$ $\la$1~mm the disk will become geometrically optically thick in the radial direction.  This further complicates the evolution and is not taken into account in our simple analysis, either in its effect on the collisional evolution, or on the brightness of the disk.

Detailed treatment of the non-axisymmetry and optical depth effects is beyond the scope of this work and will be dealt with in future papers.  Nonetheless we can broadly state that the detectable lifetimes given here for distributions with $D_{\rm{max}}$ in the strength regime ($\ll$1~km) should be an upper limit.

\subsubsection{A two component size distribution: vapour and boulders}
\label{vapandbould}

As discussed in Section \ref{sizedist}, in reality the disk is likely to be composed of two different size distributions of material, one that contains very large fragments with sizes of hundreds of kilometres and one that consists of vapour condensates.  In this case there will be a very bright initial phase during which vapour condensates are the dominant source of visible material, which will then decay rapidly.  While there are significant uncertainties in the evolution of the vapour distribution, it is certain that they would produce a very much brighter and shorter-lived signature than that of a distribution of large material extending up to hundreds of kilometres.

\begin{figure}
\includegraphics[width=85mm]{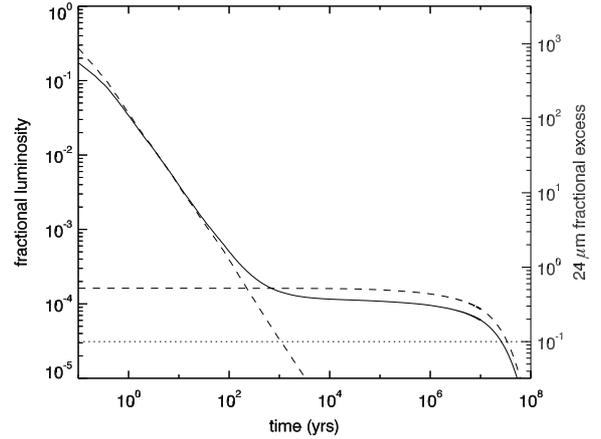}
\caption{Fractional luminosity and 24~$\mu$m excess evolution for a disk with a two component size distribution (solid line).  70\% of the mass is in a $D_{\rm{max}}=500$~km distribution and 30\% in a $D_{\rm{max}}=1$~cm distribution.  As dashed lines we show the evolution of a disk with single size distribution of $D_{\rm{max}}=500$~km or 1~cm for comparison.  The dotted line indicates a typical detection limit of $R_{24}=0.1$.}
\label{twocompevol}
\end{figure}

Once the vapour material has been ground away, we are left with dust that is replenished on much longer timescales from the large objects and so the disk enters a long-lived, lower brightness, phase.  In Fig.~\ref{twocompevol}, as an example, we show the evolution of a two component distribution, with 30\% of the mass composed of vapour condensates with $D_{\rm{max}}=1$~cm, and 70\% of the mass composed of boulder material with $D_{\rm{max}}=500$~km, consistent with the vapour fraction found by \citet{canup2008}.  The very different peak brightnesses and lifetimes of the two distributions means that the evolution is strongly divided into separate regimes where one of the distributions dominates.  Since the amount of mass in each distribution is less than the total mass of debris produced, the peak brightness in each regime is lower than that if all of the mass were contained in the dominant distribution.  This also causes the disk to become undetectable slightly faster than a single distribution with $D_{\rm{max}}=500$~km would, at 27~Myr.

\section{Comparison with known debris disks}
\label{comparison}

\begin{figure}
\includegraphics[width=85mm]{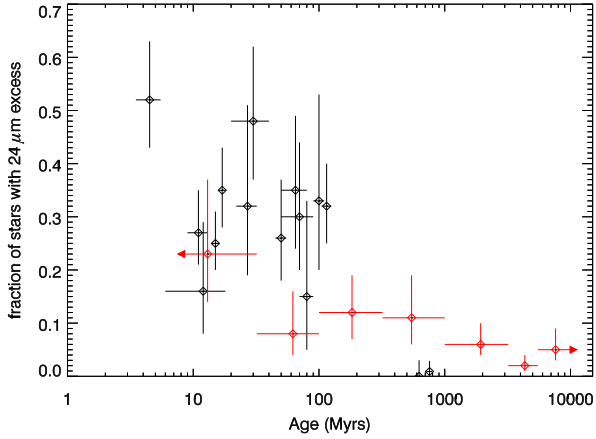}
\caption{Fractions of FGK stars with detected 24~$\mu$m excess for stars at different ages (see Table~\ref{24umtable}).  Open clusters are shown in black, for which the horizontal error bars indicate the uncertainty in the age of the cluster.  Field stars are shown in red, for which the horizontal error bars indicate the width of the bins.}
\label{24umexf}
\end{figure}

As mentioned in Section \ref{obsdisk} many surveys for debris disks around solar-type stars have used the 24~$\mu$m band of the \textsc{mips} instrument on the \emph{Spitzer} Space Telescope (e.g. \citealt{rieke2004}).  In Table~\ref{24umtable} and Fig.~\ref{24umexf} we summarise the fraction of stars with 24~$\mu$m excess that have been found for FGK stars at different ages both in open clusters and for a large sample of field stars compiled from \citet{beichman2006}, \citet{trilling2008} and \citet{carpenter2009}.  The fraction of stars displaying an excess at 24~$\mu$m falls with increasing age, from as much as 50\% at ages of $\sim$5~Myr to closer to 20\% by $\sim$100~Myr, at which time terrestrial planet formation has probably finished, and finally to only a few percent for mature stars like our own Sun.

\begin{table}
\caption{Fractions of FGK stars with 24~$\mu$m excess in open clusters of different ages and in the field.  Note that in some cases (e.g. \citealt{chen2011}) K-type stars are poorly represented.  Also note that \citet{carpenter2009b}, \citet{chen2011} and \citet{rebull2008} require F$^{\rm{obs}}_{24}/$F$^*_{24} \ge 1.3$ to formally describe an excess as detected, somewhat higher than the F$^{\rm{obs}}_{24}/$F$^*_{24} \ge$1.1-1.15 used by other papers.  The predominance of very large excesses at the ages they study (see Fig.~\ref{debriscomp}) suggests the effect on the reported excess fraction is small.  This is supported by deeper observations of a sub-set of the Sco-Cen samples in \citet{carpenter2009}.  Poisson errors in the excess fractions are determined using the tables in \citet{gehrels1986}.}
\label{24umtable}
\begin{tabular}{l @{} r@{}c@{}l r@{/}l @{} cc}
\hline
Name	&\multicolumn{3}{c}{Age (Myr)}& \multicolumn{2}{c}{Excess/total}& fraction		& ref. \\
\hline
$\lambda$-Orionis 	& 5&$\pm$&1		& 33&63 	& 0.52$^{+0.11}_{-0.09}$	& 8\\
Upper Sco		& 10&$\pm$&2		& 20&74		& 0.27$^{+0.08}_{-0.06}$	& 3, 6\\
Upper Cen Lupus		& 15&$\pm$&1		& 23&91		& 0.25$^{+0.06}_{-0.05}$	& 3, 6\\
Lower Cen Crux		& 17&$\pm$&1		& 29&82		& 0.35$^{+0.08}_{-0.07}$	& 3, 6\\
$\beta$-Pic mv. gp.	& 12&$\pm$&6		& 4&25		& 0.16$^{+0.13}_{-0.08}$	& 10\\
IC 4665$^a$		& 27&$\pm$&5		& 10&39		& 0.26$^{+0.11}_{-0.11}$	& 12\\
Tuc/Hor assoc.$^b$	& $\sim$30&&		& 19&40		& 0.48$^{+0.14}_{-0.11}$	& 15\\
IC 2391			& 50&$\pm$&5		& 6&19		& 0.32$^{+0.19}_{-0.13}$	& 9\\
NGC 2451 A,B		& 50&-&80		& 11&31		& 0.35$^{+0.14}_{-0.11}$	& 1\\
AB Dor mv. gp.		& $\sim$70&&		& 9&30		& 0.30$^{+0.14}_{-0.10}$	& 15\\
$\alpha$-Persei		& 80&$\pm$&10		& 2&13		& 0.15$^{+0.18}_{-0.10}$	& 8\\
Pleiades		& 115&$\pm$&10		& 23&71		& 0.32$^{+0.08}_{-0.07}$	& 11\\
Blanco 1		& 100&$\pm$&10		& 6&18		& 0.33$^{+0.20}_{-0.13}$	& 13\\
Hyades			& 625&$\pm$&25		& 0&67		& 0.00$^{+0.03}$		& 4\\
Praesepe		& 757&$\pm$&36		& 1&106		& 0.009$^{+0.02}_{-0.008}$	& 7\\
\hline
Field			& &-&32			& 6&26		& 0.23$^{+0.14}_{-0.09}$	& 2, 14, 5\\
			& 32&-&100		& 3&38		& 0.08$^{+0.08}_{-0.04}$	& 2, 14, 5\\
			& 100&-&320		& 6&49		& 0.12$^{+0.07}_{-0.05}$	& 2, 14, 5\\
			& 320&-&1000		& 5&45		& 0.11$^{+0.08}_{-0.05}$	& 2, 14, 5\\
			& 1000&-&3200		& 7&109		& 0.06$^{+0.04}_{-0.02}$	& 2, 14, 5\\
			& 3200&-&5500$^c$	& 3&121		& 0.02$^{+0.02}_{-0.01}$	& 2, 14, 5\\
			& 5500&-&		& 4&79		& 0.05$^{+0.04}_{-0.02}$	& 2, 14, 5\\
\hline
\end{tabular}
\emph{Refs}: (1)~\citet{balog2009}, (2)~\citet{beichman2006}, (3)~\citet{chen2011}, (4)~\citet*{cieza2008}, (5)~\citet{carpenter2009}, (6)~\citet{carpenter2009b}, (7)~\citet{gaspar2009}, (8)~\citet{hernandez2010}, (9)~\citet{siegler2007}, (10)~\citet{rebull2008}, (11)~\citet{sierchio2010}, (12)~\citet*{smith2011}, (13)~\citet{stauffer2010}, (14)~\citet{trilling2008}, (15)~\citet{zuckerman2011}. \\
$^a$ Low mass members identified by \citet{jeffries2009} only. \\
$^b$ Also includes the Columba and Carina-Near associations (see \citealt{zuckerman2011}).\\
$^c$ A few stars from \citet{beichman2006} and \citet{trilling2008} do not have age estimates, for these we assume an age equal to the mean of the other stars from the papers (4500~Myr), placing them in this bin.
\end{table}

\subsection{Putting the Moon-forming collision into context}
\label{Moondebriscomp}

\begin{figure*}
\begin{minipage}{175mm}
\includegraphics[width=175mm]{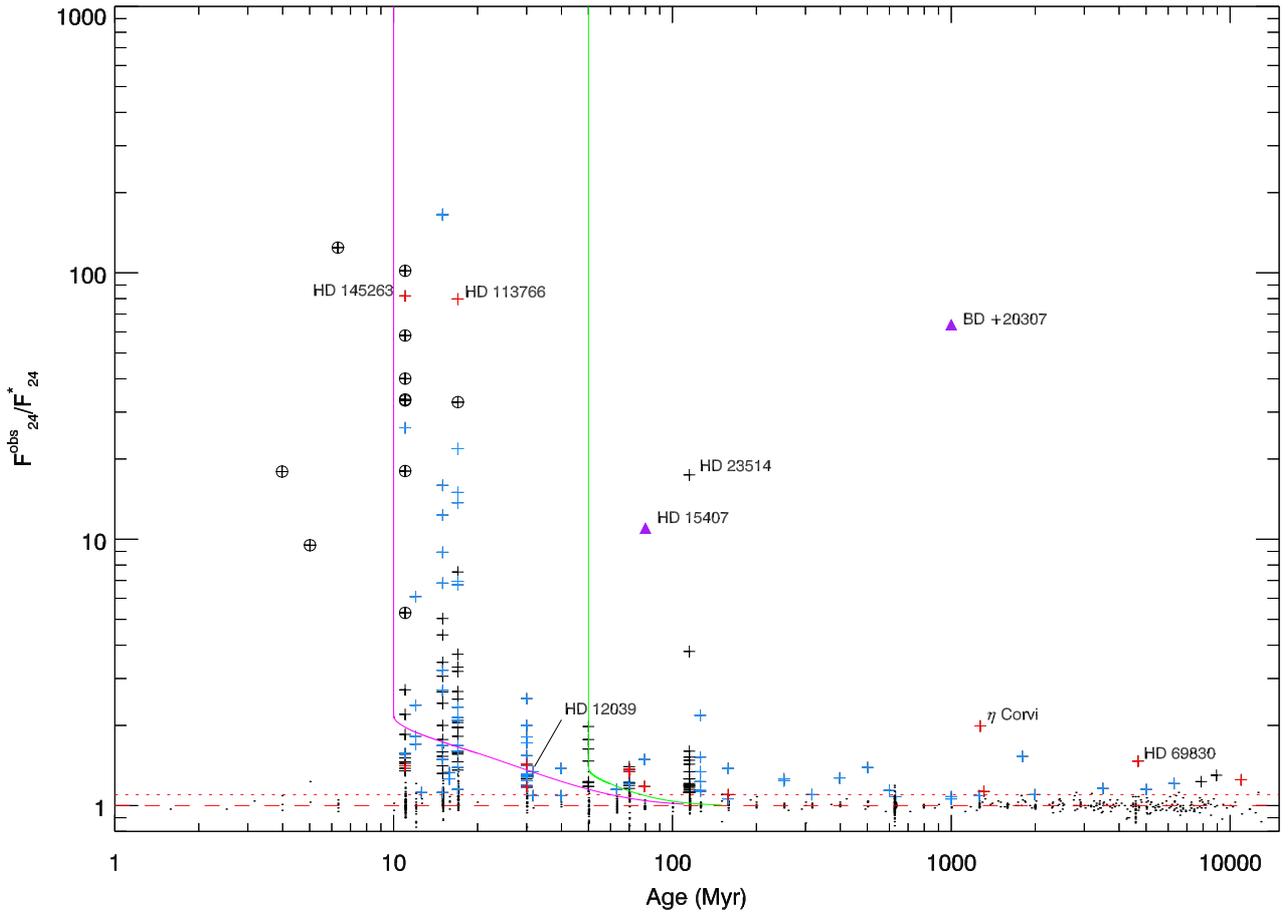}
\caption{Plot of F$^{\rm{obs}}_{24}/$F$^*_{24}$ against age for solar-type (FGK) stars from a number of different surveys at different ages and comparison with dust produced in a Moon-forming type giant impact.  F$^{\rm{obs}}_{24}/$F$^*_{24}$ is the ratio of the observed 24~$\mu$m flux to the expected 24~$\mu$m flux from the stellar photosphere, thus F$^{\rm{obs}}_{24}/$F$^*_{24}=R_{24}+1$.  The red dashed line follows F$^{\rm{obs}}_{24}/$F$^*_{24}$=1 and the red dotted line follows F$^{\rm{obs}}_{24}/$F$^*_{24}$=1.1.  F$^{\rm{obs}}_{24}/$F$^*_{24}$=1 corresponds to a purely photospheric 24~$\mu$m flux while F$^{\rm{obs}}_{24}/$F$^*_{24}$=1.1 is typically the lowest level of excess detectable for a calibration limited survey.  Detections are indicated by crosses while non-detections are indicated by dots.  Disks with dust temperatures given in the source papers (refs 2-6, 9-11, 14-15 from table \ref{24umtable}) are coloured blue for T$_{\rm{dust}}<$150~K and red for T$_{\rm{dust}}\geq$150~K, those without temperatures are black.  Circled sources are those that are considered to potentially be primordial or proto-planetary disks in the source papers.  The 24~$\mu$m excess evolution of debris from a Moon-forming type event taken from Fig.~\ref{twocompevol} is shown in green with the impact occurring at 50~Myr.  An additional, hypothetical, more massive/destructive, collision at 10~Myr is shown magenta.  Systems of particular interest are labelled and we also indicate the interesting systems BD~+20307 and HD~15407 (\citealt{weinberger2011, melis2010}) that are not included in the surveys with purple triangles.  See Section \ref{warmdustfrac} for more details.}
\label{debriscomp}
\end{minipage}
\end{figure*}

For comparison with the known debris disks we plot in Fig.~\ref{debriscomp} the evolution of the 24~$\mu$m excess for the debris from the Moon-forming impact (as given in Fig.~\ref{twocompevol} and Section~\ref{vapandbould}), with the date of impact set as 50~Myr (as is suggested by e.g. \citealt{halliday2004}, \citealt{kleine2005}).  We also, as discussed further below in Section~\ref{T}, show the evolution of a hypothetical impact at 10~Myr involving a release of 3 times the mass of debris produced by the Moon-forming impact (0.048$M_{\oplus}$ with 0.0336$M_{\oplus}$ in boulders).  We use the same debris size distributions as in Fig.~\ref{twocompevol} for this hypothetical early impact.  The debris released by this early impact is brighter and lasts for around twice as long as the Lunar formation debris.  Such an impact is well within the range of values found by \citet{stewart2012} even for a proto-Earth rather less massive than 1~$M_{\oplus}$.

The spikes produced by the small vapour condensates are rather large in comparison to the bulk of known debris disks but this phase is also rather short-lived with a lifetime of $\sim$1000~years.  These spikes can potentially explain unusual extremely luminous warm dust systems such as HD~113766, HD~145263 and BD~+20307, all of which have silicate features in their spectra (\citealt{lisse2008, honda2004, weinberger2011}), consistent with dust originating from a body of terrestrial like composition.  Although around a higher mass (A-type) star, and so not included on this plot, the infrared excess of HD~172555, with F$^{\rm{obs}}_{24}/$F$^*_{24}$=5.9 at 12~Myr, falls into the same kind of category.  The spectrum of HD~172555 also suggests the presence of large amounts of amorphous silica (distinct from silicates) and SiO gas both of which have been interpreted as indicative of a recent large hyper-velocity impact (\citealt{lisse2009}).

On the other hand the lower-level, longer-lived, component of the debris, resulting from the boulder material, compares quite well with the magnitude of more typical excess sources, which even at $\sim$10~Myr are concentrated below F$^{\rm{obs}}_{24}/$F$^*_{24}\sim$3.  Both regimes that we expect for debris resulting from giant impacts in the late stages of terrestrial planet formation are thus compatible with observations of 24~$\mu$m excesses around solar type stars.

\subsection{The fraction of stars that undergo terrestrial planet formation}
\label{TPFfrac}

As well as placing our results for debris generated by giant impact events into context with observed disks we also wish to use debris observations to investigate the fraction of stars that undergo terrestrial planet formation.  In this regard, and in particular acknowledging the relevance of this to the original Drake equation, it is convenient to cast the problem in terms of a similar equation to allow us to tease apart each of the factors that influences the predicted fraction of stars that undergo terrestrial planet formation.  Our equation is
\begin{equation}
\label{TPFeq}
F_{\rm{TPF}} = E_{24} W D T^{-1}
\end{equation}
where $F_{\rm{TPF}}$ is the fraction of stars that undergo terrestrial planet formation, $E_{24}$ is the fraction of stars that display 24~$\mu$m excesses at an appropriate age (see Section~\ref{E24}), $W$ is the fraction of those stars for which the excess is due to warm dust (see Section~\ref{warmdustfrac}), $D$ is the fraction of warm disks that are truly due to terrestrial planet formation (see Section~\ref{trulyTPFdust}), and $T$ is the fraction of the terrestrial planet formation period for which we can expect a system to display a 24~$\mu$m excess (see Section~\ref{T}).

\subsubsection{The fraction of stars with 24~$\mu$m excesses, $E_{24}$}
\label{E24}
Fig.~\ref{24umexf} provides a clear visualisation of the fraction of stars of different ages that display 24~$\mu$m excesses, but to determine $E_{24}$ we also need to define the time period we are determining it over.  As we are considering debris produced by giant impacts it is particularly the chaotic growth phase of terrestrial planet formation, in which proto-planets grow via giant impacts, that we are concerned with.

The runaway and oligarchic growth phases that precede chaotic growth through giant impacts are quite rapid with the transition to chaotic growth occurring at around, or just after, 1~Myr (\citealt{kokubo1996, kokubo1998, kenyon2006, raymond2009}) when the local density in large proto-planets exceeds that in planetesimals.  The presence of gas can stabilise the oligarchs and lengthen this phase, since it is then the combined density of the gas and planetesimals compared to that of the proto-planets that determines the transition to chaos (e.g. \citealt*{thommes2003, chambers2006}).  Gas-rich disks are comparatively short-lived however, with the observed disk fraction falling to near zero by around 6~Myr (\citealt{hernandez2007}, \citealt{mamajek2009}).  We can thus expect any system that is building terrestrial planets in this way to have transitioned to chaotic growth by, at the latest, 10~Myr and probably sooner.  The chaotic growth phase would then be expected to continue for up to around 100~Myr until the terrestrial planets reach their final configuration (e.g. \citealt{chambers2004, kenyon2006, raymond2009}).  In the case of the Solar System the Moon-forming impact at $\sim$50~Myr is generally thought to be the last giant impact to occur in the Solar terrestrial planet region.

We thus choose the time period over which we determine $E_{24}$ to be the period 10-100~Myr as this starts late enough that we can expect primordial gas disks to have fully dissipated and continues long enough to encompass the expected age range for the chaotic growth phase.  This age range is also convenient in that it covers the decade of Fig.~\ref{24umexf} within which we have most information about the fraction of stars which display 24~$\mu$m excesses.  Within this age range the 24~$\mu$m excess fractions found by individual surveys range from as little as 8\% to as much as 48\%, however they are fairly tightly concentrated around a value of roughly 30\%.  As such we find $E_{24}=0.3$ to be an appropriate value.

\subsubsection{Warm dust and terrestrial planet formation, $W$}
\label{warmdustfrac}

We have discussed the overall fraction of stars which possess 24~$\mu$m excesses in Section~\ref{E24}, but if dust is produced in terrestrial planet formation we would expect it to lie at the radial locations where the formation of rocky planets is thought to occur.  The most common division that is chosen between the inner and outer planetary system is that of the ice-line at around 150~K.  While it is possible for low mass planets lacking large gaseous envelopes to form outside the ice-line they would be of different character to those that formed within the ice-line due to the presence of large quantities of water ice.  As such the 150~K boundary is a convenient one and we define only debris disks with temperatures $\ge$150~K to contribute to $W$.

When debris disks are detected at multiple wavelengths the dust temperature is often estimated and where this is the case we can divide them up into two groups in Fig.~\ref{debriscomp}: warm disks with temperatures $\ge$150~K and cold disks with temperatures $<$150~K.  Warm disks at relatively young ages ($\la$100~Myr) are, as described above, candidates for debris produced during terrestrial planet formation, such as HD~113766 (\citealt{lisse2008}), while cold disks lie in the outer parts of the system and are more likely to be equivalents of our Kuiper belt.

Warm disks around older stars can still potentially be the result of a giant impact, as is thought to perhaps be the case for HD~69830 and BD~+20307 (e.g. \citealt{lisse2007, weinberger2011}).  In this case our previous analysis of the evolution of debris generated by giant impacts in Sections~\ref{dynamics} and \ref{collisions} is still applicable.  However it is unclear if such an impact is informative about terrestrial planet formation, given the high age, and so we restrict our analysis to stars with ages 10-100~Myr.   In the 10-100~Myr range we find that, of the 60 debris disk systems which have temperature estimates, 8 (13\%) are warm.  This suggests a value for $W$ of $\sim$0.13.

Estimating the temperature of debris disks is not an easy task, however, and there are often quite large margins of error.  One particular issue is that there may be multiple debris bands at different temperatures present in the system.  Indeed \citet{raymond2011} suggest that systems which undergo terrestrial planet formation should also possess large, cold, outer belts as the dynamically calm environments that favour terrestrial planet formation also favour retention of massive outer belts.  In this case, if a single temperature fit is used, the temperature obtained will lie somewhere within the range of the different components present.  For example $\eta$ Corvi has both cold dust at 150~AU (40~K) and hot dust at $<1.7$~AU ($\sim$370~K) (\citealt{wyatt2005}; \citealt*{smith2009}).  However \citet{beichman2006}, from whom the data for $\eta$ Corvi in Fig.~\ref{debriscomp} is taken, use a single temperature fit, obtaining 150~K, only just above our cut-off for warm dust.  $\eta$ Corvi is an unusual case, both due to its age and because both hot and cold components are particularly bright (and being nearby the cold component is resolvable and demonstrably distinct from the hot component).  Nonetheless, it serves to show that hot dust in the inner system can be concealed by cold dust in the outer system, and that high-resolution imaging can be used to break this degeneracy, as it also has been in $\eta$ Telescopii (\citealt{smith2009}).  Another method for revealing details of the location and nature of dust in debris systems is deep spectroscopy, which has been applied to systems such as HD~113766 and HD~172555.

\citet{morales2011} use a two-temperature model to study a sample of FGK-type stars which had also been studied by \citet{carpenter2009}, \citet{trilling2008} or \citet{zuckerman2011} with single temperature models. They found that a significant fraction of their sample could be better fit by a two-temperature model.  Using the \citet{morales2011} temperatures, considering the hot component where two temperatures are quoted, would almost double the number of systems with warm dust in the 20-100~Myr range to in infer $W$ to be 9/25 (36\%).

As follow-up observations of debris-bearing systems continue they will improve our constraints on $W$. In the mean time the results of \citet{morales2011} suggest that the fraction of systems with detectable 24~$\mu$m excesses that possess warm dust is unlikely to be higher than 40\%, while a conservative estimate based on the results compiled in Fig.~\ref{debriscomp} would be $\sim$13\%.  For our analysis here we will take an intermediate estimate that $W\sim$0.25.

\subsubsection{The fraction of warm dust that is truly due to terrestrial planet formation, $D$}
\label{trulyTPFdust}

While we do expect that terrestrial planet formation will result in the production of warm dust, as described above, not all warm dust need be the result of terrestrial planet formation.  Warm dust can also result from the grinding down of a massive asteroid belt analogue, or from Late-Heavy Bombardment type events scattering material in from the outer system, as suggested for the $\sim$1~Gyr old $\eta$ Corvi system by \citet{lisse2012}.  While the model of \citet{lisse2012} for the $\eta$ Corvi warm dust implies the presence of a (likely terrestrial) planetary mass object within around 3~AU, it is possible for a system to possess an asteroid belt analogue without ever having formed terrestrial planets.  As such not all of the warm dust at 10-100~Myr is necessarily the result of terrestrial planet formation and thus $D$ will be less than 1.  It is difficult to pin-down warm dust as definitely originating in terrestrial planet formation however unless high-resolution spectra are available, or the parent bodies can be found, and even then there can be some ambiguity.  Nonetheless, simply using an upper limit of 1 for the value of $D$ can still produce interesting results, though it should be borne in mind that $D$ induces a potentially significant degree of downward uncertainty in the estimate of $F_{\rm{TPF}}$.

\subsubsection{Time for which a planet forming system is expected to display a 24~$\mu$m excess, $T$}
\label{T}
We have shown in Section~\ref{collisions} that, for a reasonable size distribution of material with the majority of the debris in a distribution with a largest size of hundreds of kilometres, the debris from a Moon-forming type impact would be detectable for over 10~Myr.  Even if we reduce the size of the largest objects to around ten kilometres the debris will still be detectable for over 1 Myr, and will be brighter during that period.  The Moon-forming impact falls into the graze-and-merge category of \citet{leinhardt2011} and \citet{stewart2012}, which are the least debris producing class of giant impacts (1.6\% of the total colliding mass in the case of the Moon-forming collision).  The hit-and-run and partial accretion/erosion regimes produce debris averaging around 3\% and 5\% of the total colliding mass respectively, though there are wide variations (S. Stewart, private communication).

In Fig.~\ref{debriscomp} (in which we plot F$^{\rm{obs}}_{24}/$F$^*_{24}$ against age) we include both the Moon-forming impact and a hypothetical earlier impact at 10~Myr which produces three times as much debris (using the same debris size distribution as Fig.~\ref{twocompevol}).  This puts the hypothetical early impact at around the average debris production for a partial accretion/erosion event involving an Earth-mass planet.  Although the proto-Earth at 10~Myr is likely to have been less massive, the wide degree of variation in debris production means this is still reasonable even for a proto-Earth that is substantially less than 1~$M_{\oplus}$.  As can be seen in Fig.~\ref{debriscomp} the debris from this early impact is detectable until around 50~Myr when the Moon-forming impact occurs.  With just two impacts we have thus generated a detectable level of debris from 10 to 77~Myr ($T=0.74$).

Building terrestrial planets however is typically thought to involve not just one or two, but a whole series of giant impacts.  \citet{chambers2004, kenyon2006, obrien2006, raymond2009}, and others, all suggest that the formation of Earth and Venus requires around 10-15 giant impacts between planetary embryos each, though Mercury and Mars would require fewer mergers.  Using rather different techniques (Monte Carlo versus N-body and hydrocode), \citet{stewart2012} and \citet*{genda2011} both suggest that the whole giant impact phase of terrestrial planet formation cumulatively produces around 20\% of the final mass of the planets in debris.  If the whole terrestrial planet system is undergoing chains of giant impacts, which on average produce more debris than the Moon-forming collision, and which will be detectable for over 1~Myr, if not 10's of Myr, it is reasonable to expect a detectable level of dust for a large fraction of the giant impact phase and thus $T$ close to 1.

Though current models of terrestrial planet formation lead us to conclude that $T \sim 1$ it should be noted that should our current understanding of terrestrial planet formation change $T$ can be lower.  In particular we can think of $T$ as dependent on the number of `extra' terrestrial planets/planetary embryos that emerge from the proto-planetary disk on top of the number that constitute the final, stable, system after around 100~Myr.  Assuming that the extra embryos are removed via collisions we can then quantify the number of extra embryos and giant impacts as $g$.  If there are no extra embryos ($g=0$), and the terrestrial planets emerge from the proto-planetary disk fully formed, then $T=0$, and clearly if $T=0$, then from Eq.~\ref{TPFeq}, $F_{\rm{TPF}}$ becomes highly insensitive to $E_{24}$, $W$ and $D$, and can be arbitrarily high regardless of their values.  If there is one extra embryo ($g=1$), the minimum number that must have been present in the solar system if the Moon was formed via a giant impact, then $T$ is roughly the lifetime of the Moon-forming debris, 27~Myr, over the 90~Myr of our period of interest, so $T \sim0.3$.  If there are 2 or more extra embryos ($g\ge2$) then $T \ga 0.7$, and this is the preferred scenario based on current simulations, as indicated above.

\subsubsection{Expected number of systems in the high brightness regime}
\label{exphigh}

As we mentioned in Section~\ref{Moondebriscomp} there are two different brightness regimes we can consider and, as well as the total number of systems with detectable dust produced in giant impacts (that lie in either regime), we can also consider on the number expected to lie in the high brightness regime.  The same systems (i.e. those undergoing terrestrial planet formation) will contribute to both the total number of systems with detectable dust and those which lie in the high brightness regime.  Difference in the numbers will result solely from the difference in the length of time spent in the high brightness regime and with any detectable level of dust ($R_{24}>0.1$).  We thus define $T_h$ as the fraction of the 10-100~Myr period a system spends in the high brightness regime, as opposed to $T$ which characterises the fraction spent with any detectable level of dust.  The short amount of time spent in the high brightness regime however means that, unless impacts occur at very short intervals, $T_h$ will be directly proportional the number of giant impacts that occur during planet formation.  Taking a characteristic lifetime of the high-brightness regime as 1000~years per giant impact over 10-100~Myr period we thus have $T_h\sim 1.1\times10^{-5} g$.  As mentioned in Section~\ref{T} simulations suggest that the formation of Earth and Venus requires around 10-15 giant impacts, while Mercury and Mars would require fewer.  It can be expected that $g$ will thus fall somewhere in the region of 20-50 and so we can suggest $T_h \sim 2-5\times10^{-4}$ and probably less than $10^{-3}$.

Taking the sample of 112 systems with dust detected at 24~$\mu$m in the 10-100~Myr period there are two candidates for the high brightness regime, HD~113766 and HD~145263.  Although only 60 of the 112 systems have temperature estimates the remaining 72 mostly lie too low to be candidates for the high brightness regime.  If both HD~113766 and HD~145263 are indeed examples of systems in the high brightness regime, this leads to $T_h \sim 2\times10^{-2}$, somewhat higher than our suggested upper limit of $10^{-3}$.  Estimating a fraction based on two systems however is problematic, and this assumes that both of these systems are indeed examples of the high brightness regime.  Further making estimates of the expected value of $T_h$ is also difficult because of the non-axisymmetry for the early disk evolution, and the uncertainties for the vapour distribution.

\subsubsection{The fraction of stars that undergo terrestrial planet formation, $F_{\rm{TPF}}$}
\label{TPFfracsub}

Having discussed the factors that contribute to Eq.~\ref{TPFeq} we are now in a position to estimate $F_{\rm{TPF}}$.  Our values of $E_{24}$, $W$ and $T$ above suggest that $F_{\rm{TPF}}\la0.1$, with the downward uncertainty arising from the nature of our estimate of $D$ as an upper limit.  This indicates that terrestrial planet formation is comparatively uncommon, particularly once we take into account that $D$ is likely to be less than 1, since some young warm dust systems will be due to the grinding down of massive asteroid belts in systems that never underwent terrestrial planet formation.  Such systems can arise for example in dynamically violent environments in which eccentric and/or migrating gas giant planets disrupt the terrestrial planet formation process.  In addition, our estimated value for $W$ incorporates an assumption that a reasonable number of apparently cold dust systems also harbour warm dust, as based on the results of \citet{morales2011}.  A conservative estimate based only on dust systems that are unambiguously warm would halve our estimate to $F_{\rm{TPF}}\la0.05$.

The fraction of stars that undergo terrestrial planet formation can only be higher than our estimate if a substantially greater number of apparently cold dust systems conceal inner warm dust belts or if there is missing physics in current models of terrestrial planet formation or giant impacts.  For example if terrestrial planets were commonly fully formed by the time the protoplanetary disk dispersed, then the expected infrared excess could be much lower or even non-existent.  Alternatively $F_{\rm{TPF}}$, as inferred by this analysis, can also be raised if the debris produced by giant impacts is shorter lived or less visible such that systems are only detectable for a small fraction of the time during which they are undergoing terrestrial planet formation.  This would be possible if a substantial majority of the debris produced in giant impacts is in the form of small, millimetre-sized, dust grains, i.e. if most of the debris is produced as vapour.

We can compare the fraction of systems with terrestrial planet formation inferred from the observations of warm dust to the fractions of planet hosting systems found by planet-finding missions.  The $\eta$-Earth survey suggests a planetary occurrence rate of 12\% in the mass range 3-10~$M_{\oplus}$ (\citealt{howard2010}) on orbits shorter than 50 days, which for an appropriate mass-radius model matches well with the numbers of planetary candidates found by the \emph{Kepler} mission (\citealt{howard2011}).  It is difficult to extrapolate to lower, truly Earth-like or smaller, masses.  However, the bias-corrected occurrence rates of 5\% and 8\% found by \citet{borucki2011} for Earth-size ($1.25< R_P/R_{\oplus}$) and super-Earth-size ($1.25\leq R_P/R_{\oplus} <2$) planet candidates in a similar period range from the \emph{Kepler} mission tentatively suggest that truly Earth-like planets are no more common than super-Earths.

While this is slightly higher than our prediction of the frequency of terrestrial planet formation it is not dramatically different and there are two important points to note.  Firstly, the usual `planet occurrence rates' that are quoted are an average number of planets per star, rather than the fraction of stars that host planets.  Since there is a definite trend toward multi-planet systems (e.g. \citealt{lissauer2011, borucki2011}) this is expected to be higher than the fraction of stars that host planets, which is what is measured by $F_{\rm{TPF}}$.  Secondly, the fraction of stars that form planets in the inner parts of the system may be different to the fraction of stars that eventually host planets on relatively short orbits, due to the effects of migration.  Although the HARPS survey seems to suggest occurrence rates that are somewhat discrepant with those of the $\eta$-Earth survey and the \emph{Kepler} mission (\citealt{mayor2011}) this is primarily due to estimates of the occurrence rate in the mass range 1-3~$M_{\oplus}$ in which the detection probabilities are low and few planets are known (Pettitt et al. in prep.).  At higher masses the results of the HARPS survey are in reasonable agreement with other studies.

It should also be noted that while planet finding surveys suffer from large biases against the detection of planets with low masses and at large orbital distances dust observations provide information about the entire forming terrestrial planet system without recourse to the detection of individual planets.  This allows us to obtain information about systems in which the planets would otherwise be very difficult to detect much more easily.

\subsubsection{Developments with future instruments}
\label{futins}

Our estimates of the parameters in Eq.~\ref{TPFeq} have been based on a detection limit for debris disks at 24~$\mu$m of $R_{24}\sim0.1$.  As instrumentation improves we are likely to be able to detect debris disks to lower values of $R_{24}$.  LBTI for example is intended to be able to detect dust at 10~$\mu$m down to a level $R_{10}\sim10^{-4}$ (\citealt{hinz2008, roberge2012}).  While the debris may drop below $R_{24}\sim0.1$ at the times discussed in Section~\ref{collisions} there is still material present after this time and such a dramatic increase in sensitivity would lead to a much longer time spent above the detection threshold.  At this level even if all of the debris is produced as vapour it would still be detectable for millions of years, while the boulder distribution would generate detectable dust for hundreds-of-millions of years.  Thus even a single Moon-forming type giant impact would be sufficient to give $T=1$ for a survey with such an instrument, and moreover the age of star around which one looks for dust resulting from terrestrial planet formation can be increased.  We do caution however that at such a low level dust from terrestrial planet formation debris could be swamped by other dust sources within the host system.  One such potential source would be smaller `giant impacts' in an asteroid belt analog, dust from which may be detectable by instruments such as LBTI for a substantial fraction of a star's life.

\section{Implications for planetary compositions}
\label{compositions}

As stated in Section \ref{addVenus} (see Fig.~\ref{Nacc}), by 10 Myr after the impact Venus would have accreted a comparable amount of material to Earth, and if dynamics are the only process removing mass from the disk the amount of mass accreted would be large.  For Earth and the Moon, accretion of large amounts of material from the disk will not influence the bulk composition of the bodies since the material originated from the Earth-Moon system.  In the case of the other planets however, and in particular Venus, the amount of material accreted is particularly interesting as any material from the Earth-Moon system deposited there would potentially have a different composition from the native material.

\begin{figure}
\includegraphics[width=85mm]{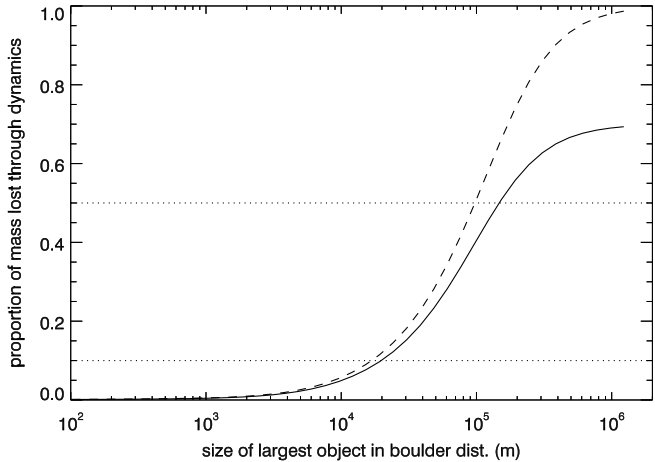}
\caption{Proportion of mass that has been lost from the disk at 10~Myr due to dynamical effects (primarily accretion) for different values of $D_{\rm{max}}$.  For the dashed line we assume that all of the mass in the disk is part of the distribution represented by the plotted $D_{\rm{max}}$ while for the solid lines we assume that 30\% of the mass was produced as vapour which is all lost to collisional grinding.}
\label{masslossdest}
\end{figure}

Exact quantities of material that can be accreted out of the disk of debris produced by the Moon-forming impact also depend strongly on how much of that material is lost through collisional evolution of the debris (as described in Section \ref{collisions}).  In Fig.~\ref{masslossdest} we show how the proportion of mass lost from the disk to collisional grinding versus dynamical losses (mostly accretion onto Earth and Venus) varies as the size of the largest object in the debris size distribution ($D_{\rm{max}}$) is varied.  If all of the initial mass of the disk is put into a single distribution (dashed line in Fig.~\ref{masslossdest}) then $>90\%$ of the mass lost at 10~Myr is lost to collisional grinding if $D_{\rm{max}}<17$~km.  On the other hand if $D_{\rm{max}}>100$~km then dynamical losses account for $>50\%$ of the mass loss and $>40\%$ of the initial disk mass is accreted onto the planets.  If 30\% of the initial disk mass is produced as vapour, which is all lost to collisional grinding (solid line in Fig.~\ref{masslossdest}), these values are revised upwards slightly to 20~km and 150~km respectively.  Thus, if the boulder distribution contains objects with sizes of several hundred kilometres, a large fraction of the mass that is lost from the disk is accreted onto the planets.  Dynamical losses dominating the mass loss from the disk also corresponds to the domination of dynamics in determining the evolution of the disk brightness.

At 10~Myr after the impact, if the disk is dynamically dominated, there is still a considerable amount of mass remaining in the disk.  So in order to know the final destination of all mass from the disk we must extrapolate beyond this.  An extrapolated version of Fig.~\ref{masslossdest} at 100~Myr rather than 10~Myr looks little different, but by this time, for any reasonable value of $D_{\rm{max}}$, a negligible amount of mass remains in the disk.  At 10~Myr 45\% of the initial dynamical particles have been lost from the VEMJ simulation, so if we assume that the ratios between the key sinks (accretion onto Earth, accretion onto Venus and ejection) remain the same until all particles have been lost the number of particles lost to each sink will increase by a factor of two.  This would place the final proportions of particles lost to accretion onto Earth, accretion onto Venus, and ejection, at 44\%, 38\% and 18\% respectively.  In the case of accretion onto Earth and Venus, the assumption that the ratio of particles lost to each remains constant is reasonable as the rates of accretion are very similar at 10~Myr.  This may underestimate the proportion ejected, however, and so overestimate the total proportion accreted, as the rate of ejection is increasing at 10~Myr (Fig.~\ref{Nacc}).  Nevertheless, uncertainties arising in the mass accreted by the different planets as a result are small in comparison to those arising from the uncertainty in $D_{\rm{max}}$.  As such in the following we will assume that the ratio of accretion onto Earth, accretion onto Venus and ejection remains constant after 10~Myr, while considering the different scenarios of 0\%, 50\% or 90\% mass loss to collisional grinding.

\subsection{Venus}
\label{Venuscomp}
As we stated in Section~\ref{addVenus}, over the 10~Myr integration time of the VEMJ simulation Venus accretes 17\% of the disk particles over the first 10~Myr.  Extrapolating out until the disk has been totally depleted Venus would then have accreted 34\% of the dynamical particles.  With no loss of mass from the disk due to collisional evolution this would amount to $3.2\times10^{22}$~kg (or 0.44 $M_{\rm{L}}$) of material, with half of this accreted during the first 10~Myr.  Although this is only 0.67\% of the mass of Venus, spread across the surface of the planet at Lunar density this would form a layer some 21 km deep.  Under a more reasonable assumption that half of the mass of the disk is lost to collisional evolution, this would fall to $1.7\times10^{22}$~kg, still enough to coat the surface of the planet to a depth of 11~km.  Even under the assumption that 90\% of the disk mass is lost to collisional evolution Venus would still have accreted a layer of material more than 2~km deep.  Depending on the degree to which this material was recycled in to the interior this raises the possibility of the upper layers of Venus containing significant amounts of material of Earth origin.  In similarity with Earth, as described below, the accretion of this material would provide a substantial source of heating to the young Venus and may be sufficient to cause disruption of the crust.  The geological history of Venus is not well known enough at present to know how well this fits with the reality, but future missions to Venus may provide indications.

\subsection{Mars}
\label{Marscomp}
Mars accretes a rather smaller amount of material than Venus (Section~\ref{addMars}).  Extrapolating out to $\sim$100~Myr, results in $\sim6\times10^{20}$~kg being accreted onto Mars (assuming no loss of material to collisional evolution).  For the more realistic assumption of 50\% of disk mass lost to collisional evolution, this would be closer to $3\times10^{20}$~kg at 100~Myr when the disk has been totally depleted, and lower still at $\sim6\times10^{19}$~kg if 90\% of the disk mass is lost to collisional grinding.  While very much less than the amount of material accreted by Venus, this is still some 3 or 4 orders of magnitude larger than the combined mass of the current Martian moons Phobos and Deimos (e.g. \citealt{andert2010, jacobson2010}) and enough to cover the surface of the planet to a depth of a few tens to a few hundred metres.

\subsection{Earth}
\label{Earthcomp}
As discussed in Section \ref{addAP} Earth accretes approximately 20\% of particles in the VEMJ simulation over the 10~Myr immediately following the Moon-forming impact.  Assuming that 50\% of the disk mass is lost through collisional grinding this amounts to roughly $10^{22}$~kg or 0.13~$M_{\rm{L}}$ during the first 10~Myr after the Moon-forming impact.  Incorporating both the potential energy released in assimilating this material and its kinetic energy assuming a typical relative velocity of $\sim$5~km~s$^{-1}$ between Earth and debris, results in accretional energy totalling around $7 \times 10^{29}$~J.  This is of similar magnitude to the excess heat contained in the magma ocean formed on Earth by the Moon-forming impact, as discussed in Section 5 of \citet{zahnle2007}.  It is also not an insignificant energy source when compared with the $\sim 3 \times 10^{30}$~J released in raising the Moon to its present orbit, both of which will be released over similar timescales.  Accretional heating from sweeping up of debris material can thus provide a significant source of heat to the early Earth and would likely prolong the magma ocean period.  In the narrative described by \citet{zahnle2007}, in which the early rate of tidal evolution of the Lunar orbit is regulated by the maximal radiation rate from a runaway greenhouse atmosphere, this would slow the rate at which the Lunar orbit widened.

We also note that the mass re-accreted by Earth from the Moon-formation debris is similar to the mass of chondritic material suggested by \citet{schlichting2012} to account for an apparent overabundance of highly siderophilic elements in Earth's crust and upper mantle (the `late veneer').  Whether or not material ejected by the Moon-forming impact could be enriched in siderophile elements is unknown, but the Moon-formation debris would certainly have interacted, and been intermingled with, any chondritic reservoir for the late veneer however.

\subsection{The Moon}
\label{Mooncomp}
As mentioned in Section~\ref{lunarinfluence} the orbit of the Moon is raised from the Roche limit, where it likely formed, quite rapidly, spending probably only 10-100~kyr within 10~$R_{\oplus}$ (\citealt{zahnle2007, schlichting2012}).  Outside $\sim$10~$R_{\oplus}$ the ratio of accretion onto Earth and the Moon, $A_{\oplus}/A_{\rm{L}}$, is relatively slowly varying, rising from 45 at 10~$R_{\oplus}$ to 60 at the present Lunar orbit of 60~$R_{\oplus}$ for impactors with a relative velocity of 5~km~s$^{-1}$ (\citealt{bandermann1973}).  An accretion ratio of $\sim$50 is thus a reasonable mean value over the 100~Myr after the Moon-forming impact.  Assuming that 50\% of the disk mass is lost through collisional grinding an accretion ratio of $\sim$50 leads to the Moon accreting $\sim4\times10^{20}$~kg of material, around 0.5\% of its present mass, by 100~Myr after the Moon-forming impact.  As described above, substantial accretion of material onto Earth may slow the rate at which the Lunar orbit is raised, leading to a lower accretion ratio at early times during which the accretion rate onto Earth is highest, and thus potentially increasing the amount of material accreted by the Moon.  As the minimum value of $A_{\oplus}/A_{\rm{L}}$ is around 25, however, this effect is unlikely to be dramatic.  Note that this accretion ratio is somewhat different to that of \citet{schlichting2012} ($\sim$200) since they use very low relative velocities for the impacting material ($\la$1~km~s$^{-1}$)

As the material accreted onto the Moon originated from the Earth-Moon system it should be of the same, or similar, composition to the material from which the Moon first accreted, and thus would probably not be traceable through compositional changes.
We do however note that the lunar late veneer complement suggested by \citet{schlichting2012} is around one quarter of the mass in Moon-formation debris we estimate to have been re-accreted by the Moon in the 50\% loss to collisional grinding scenario and that both estimates are subject to a significant degree of uncertainty.
In similarity to Earth, the accretion rate would be high, resulting in substantial deposition of energy.  The mean accretion rate during the first Myr after the impact would be $\sim 9 \times 10^{13}$~kg~yr$^{-1}$, equivalent to accretion of a 250~m impactor every day (in the 50\% loss to collisional grinding scenario), and this increases the earlier the period one considers.  This would likely disrupt the early Lunar crust and provide sufficient accretional heating to delay the solidification of the Lunar magma ocean.  This might explain some of the mis-match found by \citet{elkins2011} between the time for solidification of the Lunar magma ocean and Lunar geochronology.

\subsection{The asteroid belt}
\label{astcomp}
The debris disk produced by the Moon-forming impact extends out into the region of the present day asteroid belt (2.1-3.3~AU), although it is much less dense in this region than near the orbits of Earth and Venus (Figs.~\ref{VEMJ1Myrxy} and \ref{VEMJradhist}).  This raises the possibility that there could be some pollution of the asteroid belt by material of Earth origin.  Although early inward migration of Jupiter in the 'Grand Tack' proposed by \citet{walsh2011} would have resulted in substantial depletion of the primordial asteroid belt prior to Lunar formation, it is still expected that the belt must have been at least an order of magnitude more massive than the present $\sim6\times10^{-4}$~$M_{\oplus}$ (\citealt{krasinsky2002}) at the time of Lunar formation to account for later losses in the Late Heavy Bombardment and in 4.5 Gyrs of collisional evolution.  If we consider particles that at 10~Myr are outside Earth orbit and have eccentricities low enough that they cannot collide with any planet, which are primarily particles trapped in the Jovian 3:1 and 4:1 resonances, then for 50\% mass loss to collisional grinding this amounts to $\sim 10^{-5}~M_{\oplus}$.  This is probably an overestimate of the amount of material that can be entrained in the asteroid belt however, since particles trapped in these resonances undergo large amplitude eccentricity oscillations.  A significantly larger amount of material passes through the asteroid belt region and could undergo collisions with asteroid belt bodies, but it is unclear whether the collisional products could be retained in the belt.  Material entrained into the asteroid belt by the Moon-forming impact would thus likely have constituted only a small fraction of the mass of the belt at the time.  Modelling the dynamics involved in retaining material from the Moon-forming impact in the asteroid belt over Gyr timescales would also require the inclusion of the other outer planets to capture the detailed secular structure.

\subsection{Other giant impacts}
\label{otherGI}
Although in the case of the Moon-forming impact the masses involved, in proportion to the masses of the terrestrial planets, are not enough to induce significant changes in bulk composition, the Moon forming collision is an uncommonly efficient accretionary event.  In the giant impact outcome regimes derived by \citet{leinhardt2011}, the Moon forming collision lies within the graze-and-merge regime in which the majority of the impactor is ultimately accreted onto the target.  In contrast other giant impacts can be much more violent and as a result produce dramatically larger quantities of debris.  As an example, in their models of the formation of Mercury, \citet{anic2006} and \citet{benz2007} use an initial \emph{proto}-Mercury of chronditic silicate-iron ratio and mass 2.25 times that of the present Mercury, with thus 2/3 of the mass of the initial planet being thrown out as debris.  \citet{stewart2012} similarly found that in their simulations some impacts produced far larger quantities of debris than the average.  The single most debris producing collision was one involving a 1.26~$M_{\oplus}$ planet that produced 0.29~$M_{\oplus}$ of debris, while another 1.02~$M_{\oplus}$ planet produced as much debris over the course of its formation as its final mass.

Both \citet{stewart2012} and \citet{genda2011} find that during the process of terrestrial planet formation around 20\% of the mass of the final planets is produced in debris.  If Earth is a typical terrestrial planet then we would expect that over the course of Earth's formation $\sim0.2~M_{\oplus}$ of debris would have been produced over the course of a series of giant impacts.  While it would be expected that the distribution of kick velocities given to the debris will differ from one giant impact to another, giant impacts typically occur over a relaively small range of impact velocities relative to the escape velocity of the target, as shown in Fig.~5 of \citet{stewart2012}.  It is thus reasonable to expect that the most important variation in the distribution of kick velocities imparted to the debris will be due to differences in the escape velocity of the target, which sets the lower cut-off in the velocity distribution of the debris.  As such, leaving aside the possibility of rare high-velocity impacts, which also produce more debris, we can reasonably expect that the dynamical evolution of the debris from other giant impacts onto Earth will be similar to that of the debris from the Moon-forming impact.  As such if $\sim0.2~M_{\oplus}$ of debris is thrown off during the formation of Earth we can expect that, in the case that 50\% of the debris mass is lost to collisional grinding, Venus would have accreted $\sim$5\% of its final in Earth debris.  Even in the case that 90\% of the debris mass is lost to collisional grinding Venus would still have accreted $\sim$1\% of its final mass in Earth debris.

It has previously been recognised (e.g.~\citealt{benz2007, asphaug2010, stewart2012}) that giant impacts have the potential to lead to large changes in the bulk composition of the target/impactor.  Indeed this is one of the advantages of a giant impact in explaining the origin of Mercury.  What our results above suggest however is that, as well as having a significant impact on the composition of the target and impactor bodies themselves, giant impacts can lead to shifts in the composition of other large bodies not directly involved in the impact.  During the final phase of terrestrial planet formation it is expected that there will thus be a significant degree of indirect mixing of material across all of the terrestrial planets in a system.

\section{Conclusions}
\label{conclusions}

We have modelled the distribution of debris produced by a Moon-forming type giant impact and followed its subsequent evolution under both the dynamical influence of the other solar system planets, and through collisional evolution from self-collision.  We compare our models for the debris from Moon-formation with known debris disks and use this to provide constraints on the fraction of stars that undergo terrestrial planet formation.  We have also discussed the potential implications of the production and accretion of large masses of giant-impact-produced material for the terrestrial planets.

Dynamically we find that the initial disk is highly non-axisymmetric, and during the first few kyr after the impact this non-axisymmetry is very important in determining the rate at which material is re-accreted onto Earth.  In particular, during this phase, the vast majority of re-accretion, and non-accretionary close encounters between debris particles and Earth, occur when Earth is passing through the narrowest point in the disk.  Initially this is located at the original collision point, but gradually precesses around Earth's orbit.  The smearing out of this point as the particles precess leads to the axisymmetrisation of the disk.  Aside from Earth itself the most important bodies in determining the structure and evolution of the disk are Venus and Jupiter.  Mars plays a minor role in sculpting the disk, though accretes rather little material, while the remaining planets have very little effect.  After 10~Myr, of the initial 36,000 particles in our highest resolution dynamical simulation, 20\% have been re-accreted by Earth, 17\% have been accreted by Venus and 8\% are ejected through scattering by Jupiter while other sinks account for less than 1\%.

Collisionally we find that if, as we believe is likely, the debris contains objects with sizes of a few tens-of-kilometres or more, the dust produced by a Moon-formation like event would be detectable for millions to a few tens-of-millions of years after the impact.  If the largest objects are smaller than $\sim$100~km, collisional grinding of the debris material sets the detectable lifetime, ranging from $\sim$1000~years for distributions capped at cm-sized material, to millions of years for those including objects $\sim$10~km.  On the other hand if the largest objects $\ga$100~km, dynamical evolution becomes the dominant factor in determining the detectable lifetime, setting a maximum of about 30-40~Myr.

Current models suggest that the building of a terrestrial planet involves a chain of giant impacts, of which the Moon-forming collision is a comparatively low debris-production example.  Thus, since the debris from each impact will likely be detectable for millions, if not tens-of-millions, of years, it is reasonable to expect that a star will possess a detectable level of warm dust throughout the late stages of terrestrial planet formation. This expectation can be compared against observations of debris disks around $\sim$10-100~Myr stars at which age the final, giant impact driven, stage of terrestrial planet formation is thought to occur.  In Eq.~\ref{TPFeq} we relate the fraction of systems which undergo terrestrial planet formation to the fraction of systems that have warm infrared excesses.  We provide estimates for the fraction of 10-100~Myr stars with 24~$\mu$m excess, $E_{24}$=0.3, the fraction of those excesses that are due to warm dust, $W\sim$0.25, the fraction of that warm dust that is due to terrestrial planet formation, $D<$1, and the fraction of the 10-100~Myr period for which we expect a system undergoing terrestrial planet formation to possess a detectable 24~$\mu$m excess, $T\sim$1.  From this we conclude that the fraction of systems that undergo terrestrial planet formation, $F_{\rm{TPF}}\la 0.1$.  This is in reasonable agreement with the results of planet finding surveys such as $\eta$-Earth and \emph{Kepler}.  There is considerable downward uncertainty in $F_{\rm{TPF}}$, since $D$ is an upper limit, and there is also upward uncertainty, but a substantial increase must come with a shift in current terrestrial planet formation models.

It has been recognised that giant impacts can lead to large changes in the bulk composition of bodies directly involved in the impact.  However we suggest that giant impacts can also lead to significant indirect mixing of material, at the level of a few per cent of the mass of the terrestrial planets, through accretion of debris.

\section{Acknowledgements}
\label{acknowledgements}
APJ is supported by an STFC postgraduate studentship and MCW acknowledges the support of the European Union through ERC grant number 279973.  The authors would like to thank Sarah Stewart, Zoe Leinhardt and Grant Kennedy for valuable discussions related to the concepts presented in this paper, the anonymous referee for useful comments on the manuscript and Gillian James for conducting related preliminary work.

\bibliographystyle{mn2e}
\bibliography{refs}

\label{lastpage}
\end{document}